\documentclass{article}

\usepackage{PRIMEarxiv}

\usepackage[utf8]{inputenc} 
\usepackage[T1]{fontenc}    
\usepackage{hyperref}       
\usepackage{url}            
\usepackage{booktabs}       
\usepackage{amsfonts}       
\usepackage{nicefrac}       
\usepackage{microtype}      
\usepackage{lipsum}
\usepackage{fancyhdr}       
\usepackage{graphicx}       

\usepackage{longtable}
\usepackage{enumitem}
\usepackage{appendix}
\usepackage{color}
\usepackage{amsmath}
\usepackage{amssymb}
\usepackage{mathrsfs}
\usepackage[linesnumbered,ruled]{algorithm2e}
\usepackage{bm}
\usepackage{indentfirst}

\newtheorem{prop}{Proposition}
\newtheorem{definition}{Definition}
\newtheorem{remark}{Remark}

\title{Privacy-Preserving Adaptive Traffic Signal Control \\
in a Connected Vehicle Environment
}

\author{
  Chaopeng Tan and Kaidi Yang* \\
  Department of Civil and Environmental Engineering \\
  National University of Singapore \\
  1 Engineering Drive 2, Singapore 117576\\
  \texttt{\{tancp@nus.edu.sg, kaidi.yang@nus.edu.sg\}} \\
}

\begin{document}
\maketitle

\begin{abstract}
Although Connected Vehicles (CVs) have demonstrated tremendous potential to enhance traffic operations, they can impose privacy risks on individual travelers, e.g., leaking sensitive information about their frequently visited places, routing behavior, etc. Despite the large body of literature that devises various algorithms to exploit CV information, research on privacy-preserving traffic control is still in its infancy. In this paper, we aim to fill this research gap and propose a privacy-preserving adaptive traffic signal control method using CV data. Specifically, we leverage secure Multi-Party Computation and differential privacy to devise a privacy-preserving CV data aggregation mechanism, which can calculate key traffic quantities without any CVs having to reveal their private data. We further develop a linear optimization model for adaptive signal control based on the traffic variables obtained via the data aggregation mechanism. The proposed linear programming problem is further extended to a stochastic programming problem to explicitly handle the noises added by the differentially private mechanism. Evaluation results show that the linear optimization model preserves privacy with a marginal impact on control performance, and the stochastic programming model can significantly reduce residual queues compared to the linear programming model, with almost no increase in vehicle delay.
Overall, our methods demonstrate the feasibility of incorporating privacy-preserving mechanisms in CV-based traffic modeling and control, which guarantees both utility and privacy.
\end{abstract}


\section{Introduction}

Thanks to the recent advancements in wireless communication and sensing technologies, Connected Vehicles (CVs) have enjoyed increasing popularity in the transportation research community \cite{Feng15, Guo2019, Tan2022-optimization, Tan2022-cfd, Tan2022-est, Yang2016, deng2023cooperative}. 
Unlike traditional traffic sensors that generate aggregated data, CVs can provide detailed information about individual travelers, such as real-time trajectories, origin-destination (OD) pairs, and personal preferences. Such detailed CV data has been widely recognized as beneficial for the monitoring, control, and evaluation of urban transportation systems. However, exploiting the benefits of CVs requires collecting vehicle data at the individual level, which even if anonymized, can impose severe privacy risks on travelers, potentially leading to safety concerns or economic losses. For example, trajectory data contains highly unique human mobility patterns that can be used to infer travelers' identity, personal profile, and socioeconomic status with a high degree of accuracy~\cite{Huang2019,Li2021, Wan2018}. 
Moreover, real-time locations and speeds can reveal personal driving habits, which may be abused for discriminatory pricing in insurance~\cite{Eren2012}.  

With the imminent deployment of CVs, addressing privacy concerns in CV-based urban traffic control has become pressing. First, increasingly strict privacy protection regulations (e.g., the Personal Data Protection Act of Singapore, the General Data Protection Regulation of the European Union, etc.) impose restrictions on the collection and use of customer data, which require traffic control algorithms to be compliant with these regulations. 
Second, addressing privacy concerns is expected to incentivize more CV drivers to share their data, which can in turn improve the amount of data available to traffic controllers, thus potentially enhancing the performance of control algorithms. 

Privacy-related research in transportation mainly focuses on \emph{location privacy}, with an aim to prevent adversaries from learning (a sequence of) geographic locations of individuals. Early research either provides policy-level guidance without delving into technical details \cite{Cottrill2009, Douma2008} or designs simple anonymization and aggregation-based mechanisms \cite{Gedik2005, Kalnis2007,Beresford2003, Eckhoff2011,Liu2017, You2007,Sun2013, samarati1998protecting}. 
However, it has been widely demonstrated that anonymization and aggregation only provide a weak privacy guarantee, especially for high dimensional data, as private information can still be inferred with the help of a small external dataset   \cite{ohm2009broken}.  

More sophisticated mechanisms for protecting location privacy include two categories: differential privacy (DP)-based mechanisms and cryptography-based mechanisms \cite{kim2022privacy}. DP-based mechanisms provide a strong privacy guarantee by adding carefully designed noises to the output of query results, such that two adjacent datasets (e.g., datasets that differ in one user) produce statistically indistinguishable outputs \cite{Dewri2012, kim2021survey}. 
However, due to the added noises, DP protects location privacy at the cost of sacrificing data utility, i.e., the performance of the applications that use the shared data. In contrast, 
cryptography-based mechanisms can provide accurate query results by enabling the sharing of encrypted data without adding any noises,  which leverage cryptographic techniques including secure multi-party computation \cite{Marias,Stirbys2017}, space transformation \cite{Khoshgozaran2007}, and blockchain \cite{zhang2020decentralized}. Among these techniques, secure multi-party computation (SMPC) is a promising approach that can enable a number of participants to compute a function of their private data without having to disclose these data. However, cryptography-based techniques, including SMPC, are not resilient to inference attacks, i.e., an adversary can infer confidential information by comparing the results of several targeted queries \cite{Denning1979}.

Although these sophisticated mechanisms can help mitigate privacy risks in transportation, most existing works focus on trajectory data publication and mobile crowdsensing with minimal consideration of the characteristics of transportation systems. It is not clear how the  mechanisms proposed in these works can be applied to CV-based traffic control, which is more challenging for two reasons. First, unlike most applications where data is owned by a central trusted party, CV data is generated and stored locally on each vehicle and needs to be aggregated in a private and accurate manner to facilitate traffic control. Second, as CVs can only be gradually deployed in transportation systems, traffic control strategies need to account for low CV penetration rates in the near future. The solution to both challenges requires systematic integration of traffic modeling and privacy-preserving mechanisms to better address the tradeoff between privacy and control performance. 

To the best of our knowledge, only a few works have attempted to bridge traffic modeling/control and privacy-preserving mechanisms.   
\cite{Qin2022} devised a DP mechanism for the adaptive hyperpath choice problem to manage multimodal transportation systems while preventing the retracing of  sensitive user identities even in scenarios with data breaches. However, this work assumes the data of all users to be possessed by a central trusted third party. 
\cite{Ying2022} proposed PrivacySignal for the control of signalized intersections, in which additive secret sharing is used to construct an interactive protocol for vehicle data sharing of two roadside units (RSUs). However, this study fails to consider inference attacks and makes strong assumptions that two RSUs will not be simultaneously attacked.  
\cite{Lai2022} proposed a secure and privacy-preserving scheme for digital twin-based traffic control, which nevertheless requires a trusted third party and does not delve into traffic models. 
\cite{Tsao2022} proposed a private location-sharing mechanism for decentralized routing services. This is one of the few studies that enable the privacy-preserving optimization of transportation networks without requiring a trusted third party. However, this work requires all vehicles to be CVs and only coarsely integrates the privacy-preserving mechanisms with the Bureau of Public Roads (BPR) function without considering the modeling of traffic dynamics.

In summary, research addressing privacy concerns in traffic modeling/control is still in its infancy. We identify the following research gaps. 
First, little research on privacy-preserving traffic control can apply to scenarios without a trusted third party, which is typical for CV deployment. 
Second, despite several studies addressing privacy issues for traffic control, they focus mainly on data protection protocols without synthetically integrating traffic models. 
Third, existing research on privacy-preserving traffic modeling/control fails to address the privacy-utility tradeoff, especially the impact of DP-induced noises on control performance. 
In this study, we aim to address these research gaps by developing a privacy-preserving traffic control framework that includes (i) a privacy-preserving CV data aggregation mechanism designed for scenarios without a trusted third party and (ii) an optimization model for signal control formulated to effectively leverage the shared CV data for better privacy-utility tradeoff.

\emph{Statement of Contribution}. The contribution of this paper is three-fold. \underline{First}, we combine SMPC and DP to develop a privacy-preserving data aggregation mechanism that aggregates individual-level CV data to calculate key traffic quantities without compromising the privacy of CV users. \underline{Second}, we systematically integrate traffic models with privacy-preserving techniques by formulating traffic state estimation (e.g., arrival flow) and traffic signal control into models that can effectively utilize the outcome of the CV data aggregation mechanism. Specifically, the arrival rate estimation is formulated as a joint maximum likelihood estimation problem to mitigate the impact of the noises generated to ensure DP. The traffic signal control problem is formulated into a simple but effective linear optimization model that directly takes the outcome of the data aggregation mechanism as input. \underline{Third}, we implement a two-stage stochastic optimization model to explicitly handle the noises generated to ensure DP, which has been shown to improve the tradeoff between privacy and control performance. Overall, this study is among the pioneer works that effectively address privacy concerns in traffic signal control based on CV technology, which provides a paradigm of the privacy-preserving utilization of CV data for traffic control.

The organization of the paper is as follows. Section \ref{Section problem} introduces the problem settings and methodological framework for privacy-preserving adaptive traffic signal control. Section \ref{Section data-sharing} presents a privacy-preserving data aggregation mechanism that converts CV data to key traffic variables. Based on these variables, Section \ref{Section signal control} devises a privacy-preserving traffic signal control method. Section \ref{Section parameter} discusses the choice of critical parameters, and Section \ref{Section evaluation} presents the case studies. Section \ref{Conclusions} concludes the paper and proposes future research directions. 

\section{System Description and Methodology Framework} \label{Section problem}

\subsection{System Description} \label{subsec:system}
We consider a signalized intersection implemented with a real-time traffic signal controller based on the data provided by CVs. Although controlling an individual intersection may only require vehicles' partial trajectory information, protecting the data privacy of CVs is still important since adversaries can compromise a series of intersections or even the data center storing CV information. Even if CV data is anonymized such that their trajectories at different intersections are not linked, it is still easy to infer the routes and driving preferences of these CVs with the help of simple traffic flow models, especially in the early stage of CV deployment with a relatively low number of CVs. Moreover, the proposed algorithms for individual intersections can serve as initial building blocks that can be extended to privacy-preserving traffic control of large-scale urban transportation networks. 

We next introduce system components, as well as our assumptions on communication and adversary.  

\subsubsection{System Components}
\begin{figure}[htbp]
	\centering
		\includegraphics[scale=.2]{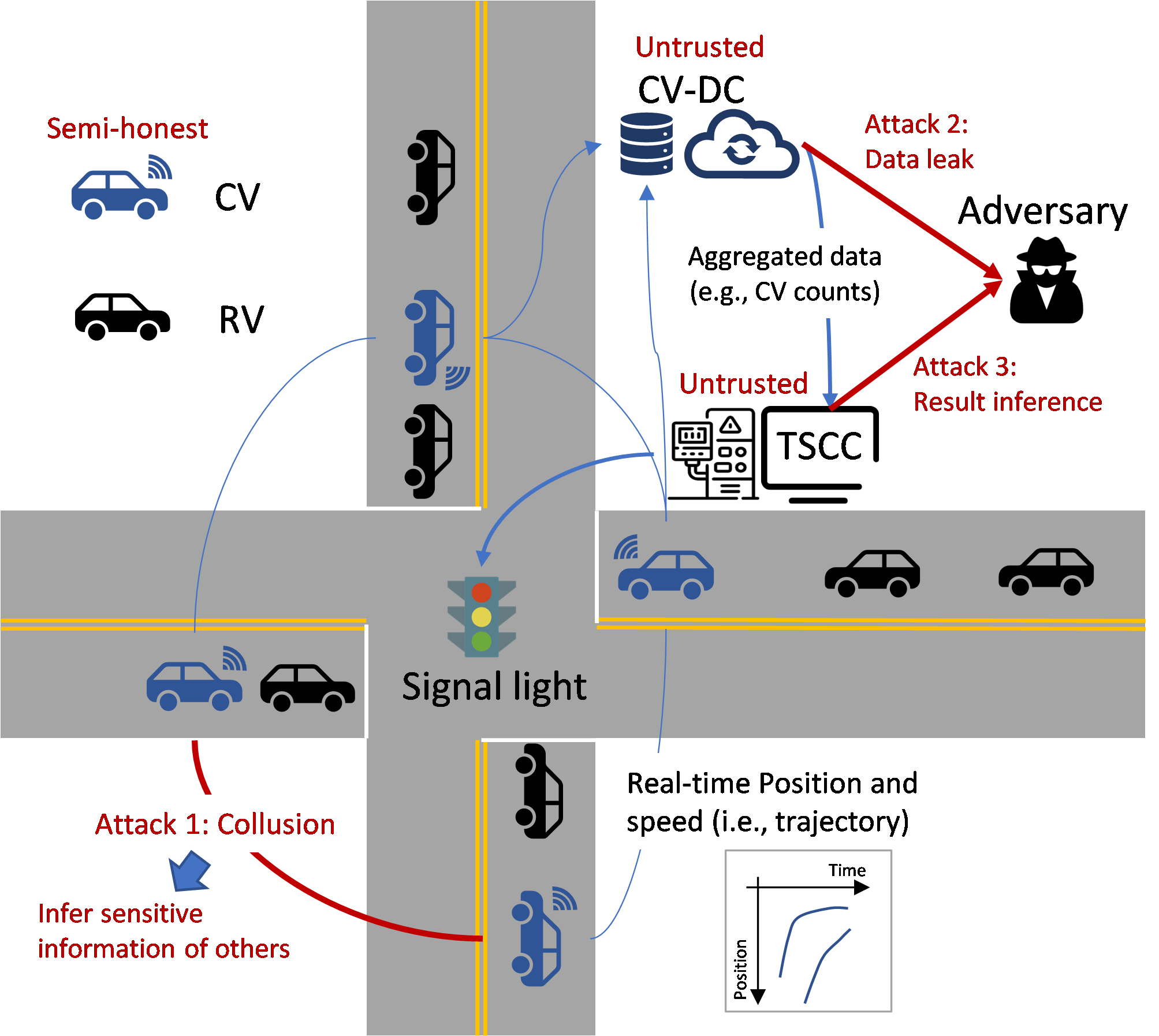}
	\caption{Illustration of CV-based traffic signal control system and adversary model.}
	\label{FIG:1}
\end{figure}

There are five components in the studied system: Connected Vehicles (CVs), Regular Vehicles (RVs), the CV Data Center (CV-DC), the Traffic Signal Control Center (TSCC), and Traffic Signal Lights (TSLs). Fig.~\ref{FIG:1} illustrates the operation of the system.\vspace{-0.3em}

\begin{itemize}[leftmargin=2em]
    \item CVs: CVs can communicate with each other and CV-DC via wireless communication. They can transmit their data (position, speed, etc.) at predetermined time intervals or in response to requests. We assume that CVs have basic capabilities of data processing (e.g., data encryption) and storage (e.g., historical trajectories). \vspace{-0.3em}
    \item RVs: RVs are not connected. Therefore, they are not involved in information transmission, and their data is completely unknown to the system.\vspace{-0.3em}
    \item CV-DC: CV-DC can aggregate and store data submitted by CVs. It enables database queries as well as the computation of basic statistics. The aggregated results will be sent to TSCC for traffic signal control. \vspace{-0.3em}
    \item TSCC: TSCC performs signal timing optimization and distributes the optimized signal timing plans to TSL based on the received data from CV-DC.\vspace{-0.3em} 
    \item TSLs: TSLs receive the signal timing plans from TSCC and execute them. TSLs will continue to execute the old plan until it receives the new plan. 
\end{itemize}

\subsubsection{Communication model} \label{Sec: Communication model}
We assume that CVs can communicate with CV-DC and other CVs within a zone of interest via encrypted wireless channels, which can be easily established using standard public key cryptography systems \cite{Weib2011}. Here, the zone of interest represents an area including directions approaching and away from the studied intersection, in which CVs can send and receive information in relation to the intersection. The radius of the zone of interest is assumed to be the minimum of the city block size and the communication range of CVs~\cite{Yang2016}. 

\subsubsection{Adversary model}
In this work, we assume that all data contributors, i.e., CVs, are honest-but-curious, meaning that they are legitimate participants who will accurately follow the communication protocol but may attempt to infer sensitive information of others from messages of intermediate results \cite{Zhao2019}.  We consider an adversary model that includes three types of attacks, as shown in Fig. \ref{FIG:1}. First, a group of CVs may collude, i.e., share their data within the group,  to infer the sensitive information of the other CVs outside the group. This could happen in scenarios where the CVs within the group belong to a single entity, such as a mobility provider. The worst case is that all but one data contributor colludes. Second, CV-DC is untrusted, which may be subject to an attack leading to data leakage in the database. Third, the TSCC may also be attacked, and the aggregated data transmitted to the TSCC by CV-DC may be intercepted to infer personal sensitive information. Note that the signal timing plan optimized by TSCC is public information and can be accessed by all parties. By securing the input data of TSCC, we can simultaneously ensure that the signal timing data does not reveal any sensitive personal information.

 \subsection{Methodological Framework} \label{Sec: Framework}

We aim to devise a privacy-preserving traffic signal control approach, which includes (i) a privacy-preserving data aggregation mechanism that calculates key traffic variables while protecting CV data and (ii) a CV-based real-time traffic signal control model that can effectively trade off privacy and control performance.  The framework of our method is shown in Fig. \ref{FIG:0}. The list of most important variables can be found in Appendix~\ref{Appendix A}. 

\begin{figure} [htbp]
	\centering
		\includegraphics[scale=.5]{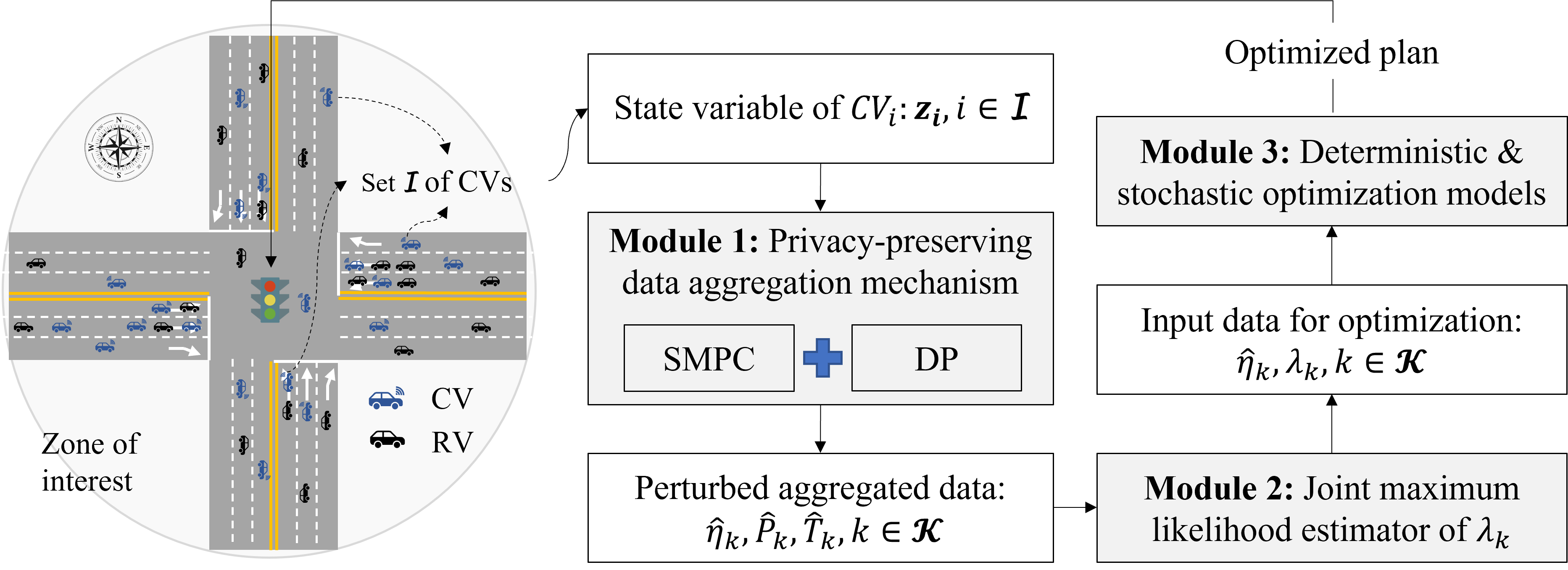}
	\caption{Methodology framework.}
	\label{FIG:0}
\end{figure}

Let $\mathcal{K}$ denote the set of all streams of the studied intersection, indexed by $k$. Let $\mathcal{I}$ denote the set of CVs in the zone of interest around the studied signalized intersection, where $N=|\mathcal{I}| \geq 2$ indicates the total number of CVs. Each CV is denoted by $ CV_i~(i\in \mathcal{I})$ with variable $\boldsymbol{z}_i$ indicating its real-time state (e.g., queuing status and position). Our proposed approach adopts a rolling-horizon optimization scheme to implement privacy-preserving CV-based traffic signal control, whereby the following modules are executed at the beginning of each decision step. \vspace{-0.3em}

\begin{itemize}[leftmargin=2em]

    \item \textbf{Module 1: Privacy-preserving data aggregation mechanism}. CVs share their data via the proposed privacy-preserving data aggregation mechanism that combines DP and SMPC. This mechanism enables CVs to collectively calculate aggregated traffic parameters (e.g., the number of CVs of each stream) without any CVs having to reveal their private data. Additionally, this mechanism also allows CVs to jointly perturb the aggregated parameters by adding carefully designed noises, ensuring even greater privacy protection. It is important to note that this mechanism does not require the involvement of a trusted third party and can preclude collusion of CVs. \vspace{-0.3em}

    \item \textbf{Module 2: Joint maximum likelihood estimator for arrival rate}. Given the aggregated parameters calculated in Module 1, CV-DC estimates the arrival rate $\lambda_k$ for each stream $k\in \mathcal{K}$ using maximum likelihood estimators, which can effectively enable robust estimation to mitigate the impact of the noises added on the aggregated parameters.  \vspace{-0.3em}

    \item \textbf{Module 3: Optimization models for signal control}. Using the aggregated parameters obtained in Module 1 and the estimated arrival rate obtained in Module 2, the TSCC solves a deterministic linear optimization model and a two-stage stochastic linear optimization model to efficiently optimize signal timings in real time. The deterministic model directly uses the input data with noises, while the stochastic model explicitly considers the distribution of noises. The optimized signal timings will be transmitted to TSLs and get executed. 
\end{itemize}

\section{Privacy-preserving Data Aggregation Mechanism for CVs} \label{Section data-sharing}
In this section, we develop our privacy-preserving data aggregation mechanism for CVs. 
Section \ref{subsec:preliminaries} introduces important preliminaries, including DP and SMPC.
Section \ref{subsec:information} presents necessary information shared by CVs for traffic signal control. 
Section \ref{subsec:mechanism} presents the privacy-preserving data aggregation mechanism based on DP and SMPC.  

\subsection{Preliminaries} \label{subsec:preliminaries}

In this section, we provide a brief introduction to DP and SMPC. Interested readers can refer to \cite{Dwork2008} and \cite{talviste2016applying} for details. 

\subsubsection{Differential privacy (DP)}
DP is a state-of-the-art notion of privacy that provides a strong and provable privacy guarantee. Specifically, let us consider the scenario where an external party (e.g., CV-DC in our case) wants to perform a query characterized by a function $\phi:\mathcal{D}\rightarrow \mathcal{X}$ that maps a dataset $D\in\mathcal{D}$ to query results $\phi(D)$ within an output space $\mathcal{X}$. 
The function $\phi$ is publicly known by both the external party and the owner of the dataset. DP protects the privacy of each data record (e.g., a CV) by ensuring that neighboring datasets, i.e., two datasets $D$ and $D'$ that differ only on this record, provide statistically indistinguishable query results.  The definition of $\epsilon$-differentially private mechanisms is given in Definition \ref{dfn:DP} \cite{Dwork2008}, where the parameter $\epsilon$ is often called the privacy budget.

\begin{definition} [$\epsilon$-Differential Privacy (DP)]\label{dfn:DP}
    Let $\mathcal{X}$ be the output space. A randomized mechanism $\mathcal{A}:\mathcal{D}\rightarrow\mathcal{X}$ satisfies $\epsilon$-differential privacy, if for any neighboring databases $D$ and $D'$ that differ on one record (e.g., adding or removing one record), and any possible output set (i.e., event) $\mathcal{E}\subset\mathcal{X}$, the following condition holds: 
    \begin{align}\label{DP}
    Pr[\mathcal{A}(D)\in \mathcal{E}]\leq e^\epsilon Pr[\mathcal{A}(D')\in \mathcal{E}].
    \end{align}
\end{definition}

For neighboring databases $D$ and $D'$ that differ on one record $\rho$,  a $\epsilon$-differentially private mechanism provides randomized responses to query $\phi$, i.e., $\mathcal{A}(D)$ and $\mathcal{A}(D')$, that follow similar probabilistic distributions with a total variation distance up to $\epsilon$. In other words, an adversary can hardly distinguish whether a given dataset is $D$ or $D'$ and thus cannot tell the existence of the record $\rho$ in the given dataset. This property enables DP mechanisms to protect against inference attacks, i.e., the limitations of most traditional mechanisms (e.g., anonymization- and aggregation-based mechanisms) in which an adversary may obtain private information by analyzing the changes in the query results, possibly with the help of external datasets \cite{Fung2010}. 

Before we introduce commonly used DP mechanisms, we need the following definition that quantifies the property of the query function $\phi$. 
\begin{definition} [Global sensitivity] \label{dfn:sensitivity}
    Let $D,D'\in \mathcal{D}$ be adjacent datasets and $\phi:\mathcal{D}\rightarrow\mathcal{X}$ be a query function. The global sensitivity of the function $\phi$, denoted by $\Delta_{\phi}$, is given as
    \begin{eqnarray}\label{GS}
    \Delta_{\phi}=\max{||\phi(D)-\phi(D')||_1},
    \end{eqnarray}
    where $||\cdot||_1$ denotes the $l_1$ norm. 
\end{definition}

 The global sensitivity defines the maximum change in the query function $\phi$ if we modify a dataset $D$ to its neighboring dataset $D'$. With Definition~\ref{dfn:sensitivity}, we present a simple yet popular mechanism to achieve DP, i.e., the Laplace perturbation (LAP) mechanism, in Proposition~\ref{prp:LAP} \cite{Dwork2008}. 

\begin{prop} [Laplace perturbation (LAP) mechanism] \label{prp:LAP}
    For query function $\phi:\mathcal{D}\rightarrow\mathbb{R}$,  a Laplace perturbation (LAP) mechanism $\mathcal{A}_{\phi}$ satisfies $\epsilon$-differential privacy. Here, the LAP mechanism $\mathcal{A}_{\phi}$ is defined as a mechanism that adds independently generated noises $\xi \sim Lap(0,b)$ to the query results, i.e., 
    \begin{align}\label{LAP}
    & \hat{\phi}(D)=\phi(D)+\xi,
    \end{align}
where the probability density function is $f_{\xi}(v)=\frac{1}{2b}e^{-\frac{|v|}{b}}$, $b=\Delta_{\phi} / \epsilon$, and $\Delta_{\phi}$ indicates the global sensitivity of the function $\phi$.
\end{prop}

We make the following remarks regarding the LAP mechanism. First, the introduction of Laplace noises can affect the accuracy of the query results, which can in turn influence the performance of the applications that use the query results (i.e., data utility). Second, LAP can be implemented as both global DP and local DP. Global DP applies to scenarios where a trusted third party can collect the data from all users and provide randomized responses to queries based on its collected dataset. However, global DP is not suitable for our case since our third party, i.e., CV-DC, is susceptible to privacy attacks and thus may not be trusted. 
In contrast, local DP applies to scenarios without a trusted third party, whereby individual users perturb their data before sharing. However, it is evident that local DP may lead to more perturbations on the aggregated results compared to global DP and hence may further deteriorate data utility. Therefore, local DP is not desirable for our case either. To address the privacy-utility tradeoff, we will integrate global DP with the promising SMPC that can help relax the requirement of a trusted third party. 

\subsubsection{Secure multi-party computation via additive secret sharing} \label{subsubsec:SMPC}
Secure multi-party computation (SMPC) is adopted to relax the dependence on a trusted third party. SMPC addresses the question of how to jointly and securely compute a convention function (e.g., a polynomial) without a trusted third party, i.e., keeping the data of contributors private \cite{Zhao2019}. 

We implement SMPC via a simple method based on additive secret sharing \cite{talviste2016applying}. The reason for adopting additive secret sharing is that it ensures input privacy with information-theoretic security and correctness of data aggregation, which means it can maintain security against adversaries with unbounded computing resources and time while guaranteeing correctness. 

The basic idea of additive secret sharing is as follows. Suppose there are $N$ agents denoted by $a_1,a_2,...,a_N$, and one agent wants to share a secret value $x$ with the others. Instead of directly disclosing the value of $x$, this agent creates $N$ shares $s_1,s_2,...,s_{N}$ out of $x$ such that $x=\sum_{i=1}^N s_i$ mod $p$ and send share $s_i$ to agent $i$. Specifically, given a large prime integer $p$, shares $s_1,s_2,...,s_{N-1}$ are chosen independently and uniformly at random from the set $\{0,1,2,...,p-1\}$, and then $s_N$ is determined by $s_N=x-\sum_{i=1}^{N-1}s_i$ mod $p$. To reconstruct $x$, $N$ agents can simply add all shares modulo $p$ together since we have $x=\sum_{i=1}^{N}s_i$ mod $p$ (from now on, we omit the mod $p$ notation for brevity). 

Note that additive secret sharing can withstand any collusion of up to $N-1$  agents since they have random values that do not depend on $x$. However, there is no guarantee of DP in the final results. An adversary can still infer private personal information from the query results.

In the rest of this section, we will introduce the data provided by CVs and then combine DP and SMPC to aggregate these data. 

\subsection{Information shared by CVs} \label{subsec:information}
Unlike existing CV-based signal control methods that require detailed CV trajectories, we only request CVs to share, at the beginning of each decision step, the minimum amount of information necessary for traffic signal control.

\begin{figure}[htbp]
	\centering
		\includegraphics[scale=.4]{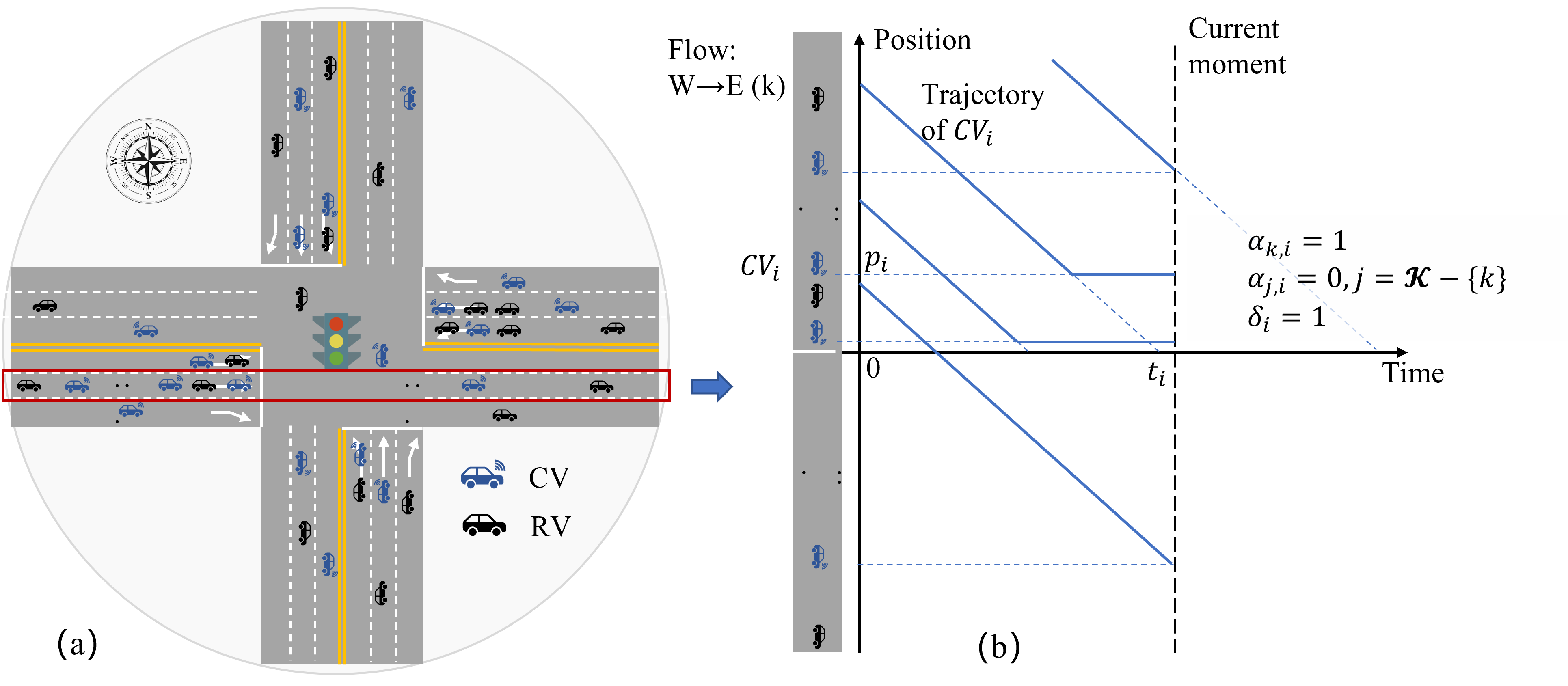}
	\caption{CVs participating in data aggregation: (a) CVs at the studied intersection; (b) state variables of a CV.}
	\label{FIG:2}
\end{figure}

As shown in Fig. \ref{FIG:2}(b), the state of $CV_i$ ($i\in \mathcal{I}$) at each decision step can be characterized by  $\boldsymbol{z}_i = \Big( \{\alpha_{k,i}\}_{k\in \mathcal{K}}, \delta_i, p_i, t_i\Big)$, whereby these variables are defined as follows.\vspace{-0.3em} 
\begin{itemize}[leftmargin=2em]
    \item $\alpha_{k,i}$ indicates whether $CV_i$ is in stream $k$. $\alpha_{k,i}=1$ if $CV_i$ is in stream $k$ and 0 otherwise. $CV_i$ generates a vector $\{\alpha_{k,i}\}_{k\in \mathcal{K}}$ including its presence in all streams satisfying $\sum_{k\in \mathcal{K}}\alpha_{k,i} = 1$, i.e., only one element is 1 since $CV_i$ can only be present in one stream. Such treatment allows the CV to participate in the calculation of traffic variables of all streams, which can help hide the stream to which the CV belongs when leveraging the proposed privacy-preserving data aggregation mechanism in Section ~\ref{subsec:mechanism}.\vspace{-0.3em} 
    \item $\delta_i$ indicates whether $CV_i$ is queued before the stopline, where $\delta_i=1$ indicates that $CV_i$ is queued (i.e., its speed is lower than a threshold, e.g., 5km/hr) and 0 otherwise.  \vspace{-0.3em} 
    \item $p_i$ indicates the current position of $CV_i$ (in terms of the number of vehicles) measured from the stopline, which can be calculated as $p_i=L_i/L_0$, where $L_i$ represents the distance to the stopline, and $L_0$ represents the jam spacing for queued vehicles. Without loss of generality, we convert $L_i$ of all CVs to $p_i$, where $p_i$ can be a negative value if a CV has already passed through the stopline. We can later use $\delta_i$ to filter out vehicle $i\in\mathcal{I}$ with a negative $p_i$.\vspace{-0.3em} 
    \item $t_i$ indicates the (expected) arrival time to the stopline of $CV_i$ [s] (relative to the start time of red signal). If $CV_i$ has not passed through the stopline (either queued or not), $t_i$ is calculated as the virtual arrival time, i.e., the time of arrival if $CV_i$ had not joined the queue. Otherwise, $t_i$ is the actual passing-through time of the CV (this time can be recorded by CVs).   \vspace{-0.3em} 
\end{itemize}

Given the above state variables, each CV prepares the following private data: counting variable $\eta_{k,i}=\alpha_{k,i}\delta_i$, position variable $P_{k,i}=\alpha_{k,i} \delta_i p_i$, and arrival-time variable $T_{k,i}=\alpha_{k,i} \delta_i t_i$ ($i\in \mathcal{I}$ and $k \in \mathcal{K}$). The prepared data can be aggregated using the privacy-preserving data aggregation mechanism to derive variables describing traffic condition in each stream, which will be further used for arrival rate estimation (in Section \ref{Sec: JMLE}) and traffic signal control (in Section \ref{Sec: deterministic model} and \ref{stochastic model}). 
For example, the total number of queued vehicles in stream $k$ can be derived by aggregating the counting variables, i.e., $\sum_{i\in\mathcal{I}}{\eta_{k,i}}$. This is because $CV_i$ is in stream $k$ if and only if $\alpha_{k,i}=1$ and queued before the stopline if and only if $\delta_i=1$. For CVs with $\eta_{k,i}=0$, their participation in the calculation does not change the result.

\subsection{Privacy-preserving data aggregation mechanism} \label{subsec:mechanism}
In this section, we develop a privacy-preserving data aggregation mechanism (i.e., Step 1 in Fig.~\ref{FIG:0}) in the absence of a trusted third party, which combines SMPC and DP to calculate the sum of private values owned by individual CVs without any CV having to disclose its private value. 

\begin{figure} [htbp]
	\centering
		\includegraphics[scale=0.7]{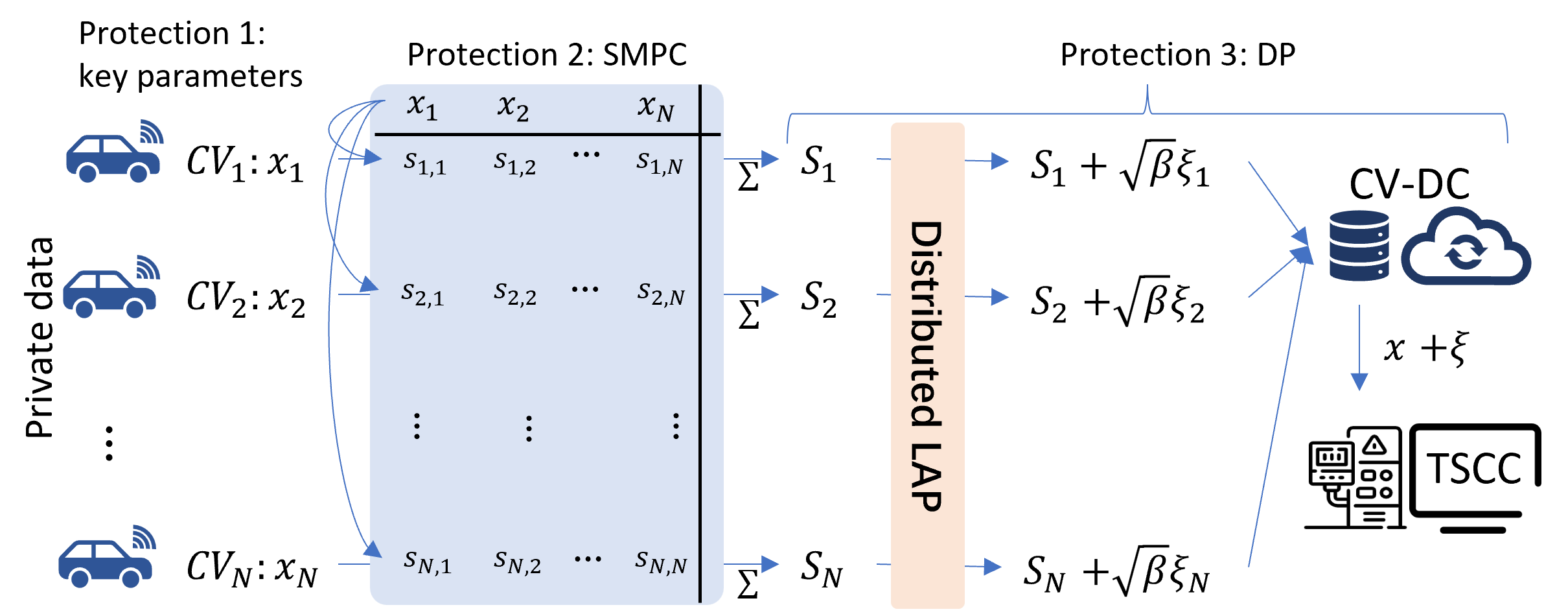}
	\caption{Illustration of privacy-preserving data aggregation of $N$ CVs by SMPC and DP.}
	\label{FIG:3}
\end{figure}

We use $x_i \in \{\eta_{k,i}, P_{k,i}, T_{k,i} \}_{k\in \mathcal{K}}$ to denote the private data of $CV_i$. 
CVs within the zone of interest want to jointly compute the sum $x=\sum_{i \in \mathcal{I}}{x_i}$ without having to disclose their private value $x_i,~i\in \mathcal{I}$. The implementation process of the proposed mechanism is shown in Fig. \ref{FIG:3}.  For brevity, we require each CV to exchange messages with all other CVs within the zone of interest. Nevertheless, this requirement can be relaxed to allow $CV_i,~i\in \mathcal{I}$ to exchange messages with CVs in a different set $\mathcal{I}_i$, e.g., with CVs within a certain radius of $CV_i$. To do so, interested readers can simply replace $\mathcal{I}$ with $\mathcal{I}_i$ in the subsequent equations for $CV_i,~i\in \mathcal{I}$ whenever applicable.

The mechanism includes triple protections to preserve the privacy of individual CV data: \vspace{-0.3em}

\begin{itemize}[leftmargin=2em]
    \item \textbf{Protection 1: Key parameter extraction.} To reduce the amount of shared data, we only require CVs to share key parameters (i.e., Section \ref{subsec:information}) rather than the complete trajectories. Thereby, once the key parameters are protected, all information about the trajectories, such as the expected stream direction, queuing status, route information, and ODs cannot be inferred.  \vspace{-0.3em}

    \item \textbf{Protection 2: Private and secure data aggregation via SMPC.} 
We employ SMPC to aggregate CV data in a private and secure manner. Specifically, $CV_i$ leverages additive secret sharing and divides its private value $x_i$ to $|\mathcal{I}|$ shares of random values $\{s_{ij}\}_{j \in \mathcal{I}}$ such that $x_i=\sum_{j \in \mathcal{I}} s_{ij}$, where the $j$-th share will be sent to $CV_j$. $CV_j$, upon receiving all shares $\{s_{i'j}\}_{i' \in \mathcal{I}}$ from other CVs, will calculate and submit $S_j=\sum_{i' \in \mathcal{I}} s_{i'j}$ to CV-DC. We make two remarks for this procedure. First, we note that $ \sum_{i \in \mathcal{I}}{x_i} = \sum_{i \in \mathcal{I}}\sum_{j \in \mathcal{I}} s_{ij} = \sum_{i \in \mathcal{I}}{S_i}$.  Second, as described in Section~\ref{subsubsec:SMPC}, this procedure can withstand any collusion of up to $N-1$ CVs. In other words, as long as each CV keeps its own share private, the other CVs will not be able to infer its private value. \vspace{-0.3em}

\item \textbf{Protection 3: DP guarantee.} We apply DP to further perturb the aggregated results to protect against inference attacks. This is especially important if the number of CVs in set $\mathcal{I}$ is small. Note that the perturbation should be applied  on $S_j$ before it is submitted to CV-DC. Otherwise, untrusted CV-DC will have access to the raw data of the unperturbed aggregated values, which could pose privacy risks.  
Therefore, unlike the traditional DP that directly perturbs the query results with the LAP mechanism, here we use a distributed LAP mechanism that perturbs the shared data right at the CV side while guaranteeing DP at the same level. In this way, the personal data stored in CV-DC is already encrypted by SMPC and perturbed by distributed LAP. Even if an adversary has direct access to the database, the sensitive information of individuals cannot be obtained by inference attacks.
Proposition \ref{prp:DLAP} presents one possible distributed LAP mechanism that applies to each portion of data, which can ensure that the aggregation of these portions satisfies $\epsilon$-differential privacy \cite{Goryczka2015}. 
\end{itemize}

\begin{prop} [Distributed LAP mechanism] \label{prp:DLAP}
    Given a random variable $\beta \sim Beta(1, N-1)$ and independent and identically distributed random variables $\xi_i \overset{i.i.d}{\sim} Lap(0,b)$, random variable $\xi=\sqrt{\beta}\sum^N_{i=1}\xi_i$ follows Laplace distribution $Lap(0,b)$. 
\end{prop}

By Proposition~\ref{prp:DLAP}, instead of directly submitting $S_i$, $CV_i$ will submit perturbed data $\hat{S}_i=S_i+\sqrt{\beta}\xi_i$ to CV-DC. Such a treatment can ensure the aggregated variable reconstructed by CV-DC satisfies $\epsilon$-DP, i.e., 
\begin{align} \sum_{i \in \mathcal{I}}\hat{S}_i=\sum_{i \in \mathcal{I}}\Big(S_i+\sqrt{\beta}\xi_i\Big)=\sum_{i \in \mathcal{I}}S_i+\sum_{i \in \mathcal{I}}\sqrt{\beta}\xi_i = \sum_{i \in \mathcal{I}}{x_i} + \xi,
\end{align}
where random variable $\xi=\sum_{i \in \mathcal{I}}\sqrt{\beta}\xi_i \sim Lap(0,b)$.  

In summary, the proposed privacy-preserving data aggregation mechanism is presented in Algorithm~\ref{alg: privacy mechanism}. 

\begin{algorithm} [htbp] 
\caption{Algorithm for privacy-preserving data aggregation mechanism} \label{alg: privacy mechanism}
\SetKwInOut{Input}{Input}
\SetKwInOut{Output}{Output}
\Input{private data $\{x_{i}\}_{i \in \mathcal{I}}$ ($ |\mathcal{I}| \geq 2$), a large prime value $p$, sensitivity $\Delta^x$, and privacy budget $\epsilon^x$}
\Output{aggregated data $\hat{x}$ with perturbations and $\hat{x} = \sum_{i \in \mathcal{I}}{x_{i}} +\xi^x$}
\For {$i \in \mathcal{I}$}
{
    \For {$j \neq i \in \mathcal{I}$}
    {
    $CV_i$ uniformly and randomly samples $s_{ij}$ from $ \{1,2,...,p-1\}$\;
    $CV_i$ sends data $s_{ij}$ to $CV_j$\;
    }
    $CV_i$ calculates $s_{ii} = x_{i} - \sum_{j \neq i \in \mathcal{I}} s_{ij} \mod{p}$\;
}
CV-DC generates $\beta\sim Beta(1,|\mathcal{I}|-1)$ and sends it to all CVs\;
\For {$j \in \mathcal{I}$}
{
    $CV_j$ generates random variable $\xi_j^x \sim Lap(0,\Delta^x / \epsilon^x)$\;
    $CV_j$ calculates $S_{j} =  \sum_{i \in \mathcal{I}} s_{ij}$ \;
    $CV_j$ calculates $\hat{S}_j = S_j + \sqrt{\beta} \xi_j^x$ and sends it to CV-DC \;
}
CV-DC calculates $\hat{x}=\sum_{i \in \mathcal{I}} \hat{S}_i$,  
\Return $\hat{x}$
\end{algorithm}

\section{Privacy-Preserving Traffic Signal Control}  \label{Section signal control}
\begin{figure}[htbp]
	\centering
		\includegraphics[scale=.6]{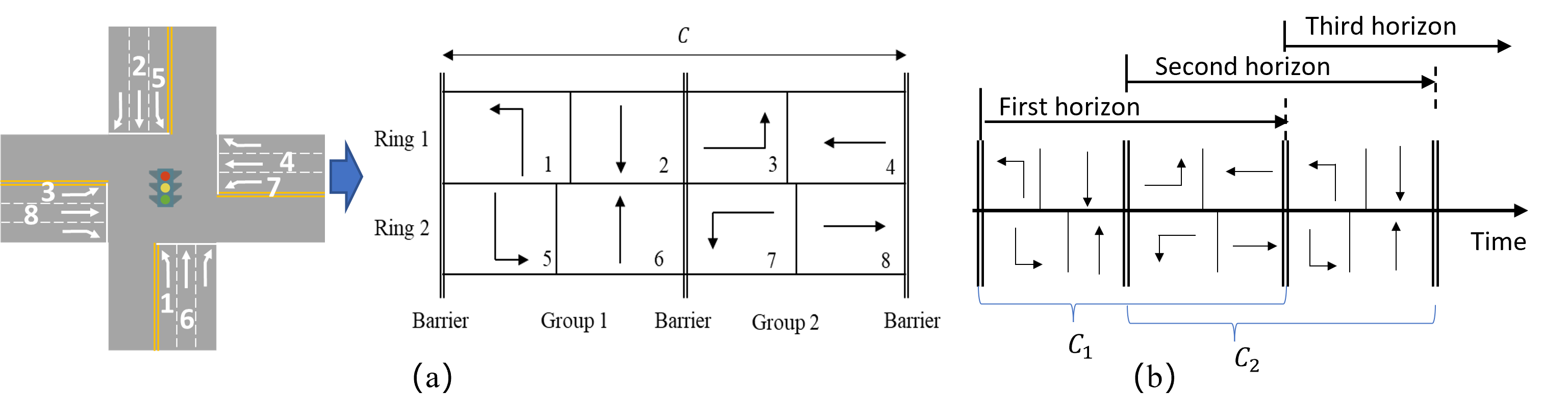}
	\caption{Basic settings of traffic signal optimization: (a) example phase sequence; (b) rolling optimization scheme}
	\label{FIG:4}
\end{figure}

In this section, we devise privacy-preserving traffic signal control algorithms based on the output of the data aggregation mechanism proposed in Section~\ref{Section data-sharing}. For presentation convenience, we focus our modeling on a NEMA ring-barrier structured intersection with uncontrolled right-turning maneuvers, as illustrated in Fig. \ref{FIG:4}(a). Note that the proposed method can be easily extended to more general intersections by modifying only a small set of constraints. 

We adopt a rolling-horizon optimization scheme for real-time signal control, as shown in Fig. \ref{FIG:4}(b). Unlike most approaches that make decisions at the beginning of each cycle, we solve an optimization problem at the end of each phase group (i.e., every half cycle) to obtain the optimal signal timings of the entire cycle. Here, we increase the decision-making frequency because the vehicle arrival information collected from CVs at isolated intersections is limited by the length of links and hence can provide a myopic view, which requires more frequent re-optimization based on updated observations, especially in scenarios where traffic fluctuates significantly. 

Before we formulate the signal timing optimization model, we present an estimator for the arrival rate in Section~\ref{Sec: JMLE}, which is a critical input for signal timing optimization models. Then, we will formulate a deterministic optimization model in Section \ref{deterministic model} and extends it to a stochastic optimization model to account for the parameter perturbations generated to ensure DP in Section \ref{stochastic model}. 

\subsection{Joint maximum likelihood estimator of arrival rate} \label{Sec: JMLE}
The key parameters extracted in Section~\ref{subsec:information} can be used to reconstruct arrival rates, which serve as important parameters for traffic signal control. Existing research on CV-based arrival rate estimation (e.g., \cite{yang2018queue}) tends to leverage detailed trajectory information, which over-mines the trajectories without considering privacy issues. Instead, we will use only the perturbed aggregated traffic variables calculated in the privacy-preserving data aggregation mechanism to estimate arrival rates. To this end, we employ a 
joint estimation method for multi-stream arrival rates proposed by \cite{Tan2022-est}, which can directly take the aggregated traffic variables as input and achieve satisfactory accuracy even in scenarios with low CV penetration rates.

For each cycle, stream $k \in \mathcal{K}$ has observed a set of queued CVs $\mathcal{I}_k^{\rm{q}}$ during the red time. 
Let $\gamma_k$ denote the proportion of CVs in stream $k$ out of the CVs in all streams over a historical period. Then, the arrival rate of stream $k$ during the corresponding red time can be estimated by the following estimator that maximizes the joint likelihood function of multiple streams (the derivation process can be seen in Appendix~\ref{Appendix B}): 
\begin{eqnarray}\label{lambda estimator}
\lambda_k= \gamma_k \frac{\sum_{k' \in \mathcal{K}} \sum_{i \in \mathcal{I}_{k'}^{\rm{q}}} p_i}{\sum_{k' \in \mathcal{K}} \gamma_{k'} \sum_{i \in \mathcal{I}_{k'}^{\rm{q}}} t_i}.  
\end{eqnarray}
Eq.(\ref{lambda estimator}) is derived by approximating vehicle arrival as a Poisson process, which is accurate for isolated intersections where each stream is undersaturated. Note that we do not expect such approximation to cause significant errors in reality. This is because it is usually possible to determine signal timings such that the intersection is not heavily oversaturated for all streams, which can also be desirable due to the existence of the lost time. 

In our cases, the proportion of CVs $\gamma_k$ is calculated based on the number of queued CVs during the past $c$ cycles, i.e., 
    \begin{eqnarray}\label{alpha calculation}
    \gamma_k=\frac{\sum_{r=1}^c N_{k,r}^{\rm{q}}}{\sum_{r=1}^c \sum_{k' \in \mathcal{K}} N_{k',r}^{\rm{q}}},
    \end{eqnarray}
where $N_{k,r}^{\rm{q}}$ indicates the number of queued CVs in stream $k$ during the past $r$-th cycle ($r=1$ indicates the current cycle); hereafter, the subscript/superscript $r$ indicates the corresponding variable in the past $r$-th cycle.

As is shown, the arrival rate estimator only requires the aggregated data of queued CVs at each stream, i.e., $N_{k,r}^{\rm{q}}$, $\sum_{i \in \mathcal{I}_{k}^{\rm{q}}} p_i$, and $\sum_{i \in \mathcal{I}_{k}^{\rm{q}}} t_i$, which can be effectively protected by the proposed privacy-preserving data aggregation mechanism. 

Using $\sum_{i \in \mathcal{I}_{k}^{\rm{q}}} p_i$ as an example, by having all CVs at the studied intersection participate in the calculation for each stream, we can easily obtain that 
\begin{align}
    & \sum_{i\in \mathcal{I}_{k}^{\rm{q}}} p_i=\sum_{i\in \mathcal{I}_{k}^{\rm{q}}} P_{k,i}+\sum_{i\not \in \mathcal{I}_{k}^{\rm{q}}} P_{k,i}=\sum_{i\in \mathcal{I}} P_{k,i} \triangleq P_k, \label{SumP} 
\end{align}
where $P_{k,i}=p_i$ for $i\in \mathcal{I}_{k}^{\rm{q}}$; $P_{k,i}=0$ for $i\not \in \mathcal{I}_{k}^{\rm{q}}$.

Similarly, we have $N_{k,r}^{\rm{q}}=\sum_{i\in \mathcal{I}} \eta_{k,i}^r \triangleq \eta_k^r$ and $\sum_{i\in \mathcal{I}_{k}^{\rm{q}}} t_i=\sum_{i\in \mathcal{I}} T_{k,i}\triangleq T_k$, where we use $\eta_{k,i}^r$ to represent the counting variable of the past $r$-th cycle. 
Then, the estimator can be rewritten as:
\begin{align}\label{lambda estimator 1}
    & \lambda_k= \gamma_k \frac{ \sum_{k' \in \mathcal{K}} P_{k'}}{\sum_{k' \in \mathcal{K}} \gamma_{k'} T_{k'}}
 \quad \text{where} \quad \gamma_k = \frac{\sum_{r=1}^c \eta_{k}^r}{\sum_{r=1}^c \sum_{k' \in \mathcal{K}}\eta_{k'}^r}.
\end{align}

Note that Eq. (\ref{lambda estimator 1}) is an estimate of the arrival rate with accurate aggregated data from CVs. After we apply the proposed privacy-preserving data aggregation mechanism, we have that each CV submits perturbed data $\hat{S}_{k,i}^x$ rather than private data $x_{k,i}, (x \in \{ \eta, P, T \})$. Revisiting Section \ref{subsec:mechanism}, we have
\begin{align}
    & \hat{S}_{k,i}^x=S_{k,i}^x+\sqrt{\beta_k^x} \xi_{k,i}^x,  \label{add noise}
\end{align}
where $\hat{S}_{k,i}^x$ is the encrypted value shared by $CV_i$; $S_{k,i}^x$ is the corresponding share based on SMPC; $\xi_{k,i}^x$ is the corresponding random noise generated by distributed LAP mechanism based on DP, i.e., $\xi_{k,i}^x \sim Lap(0,\Delta_k^x / \epsilon_k^x)$; $\beta_k^x$ is a random value following Beta distribution, i.e., $\beta_k^x \sim Beta(1,N-1)$, where $N$ is the total number of CVs at the studied intersection.

Then, by aggregating all values shared from CVs, we have 
\begin{align}
    & \hat{P}_k=\sum_{i \in \mathcal{I}} \hat{S}_{k,i}^P=\sum_{i \in \mathcal{I}} S_{k,i}^P + \sqrt{\beta_k^P} \sum_{i \in \mathcal{I}} \xi_{K,i}^P =P_k+\xi_k^P, \label{FinalP}
\end{align}
where $\xi_k^P$ is Laplace noises guaranteeing $\epsilon$-DP; $\hat{P}_k$ is the encrypted value sent to TSCC for traffic signal control if both SMPC and DP are used.

Similarly, we also have 
 $\hat{\eta}_k=\eta_k+\xi_k^\eta$ and $ \hat{T}_k=T_k+\xi_k^T$. 
Thereby, the arrival rate estimator with perturbed aggregated data is written as
\begin{align}\label{lambda estimator 2}
    & \lambda_k= \gamma_k \frac{\sum_{k' \in \mathcal{K}} \hat{P}_{k'}}{\sum_{k' \in \mathcal{K}} \gamma_{k'} \hat{T}_{k'}} \quad \text{where} \quad  \gamma_k = \frac{\sum_{r=1}^c \hat{\eta}_{k,r}}{\sum_{r=1}^c \sum_{k' \in \mathcal{K}} \hat{\eta}_{k',r}}.
\end{align}

\subsection{Deterministic optimization model} \label{Sec: deterministic model}
\label{deterministic model}
In this section, we formulate CV-based traffic signal control into an optimization model. This model is parameterized by the aggregated traffic variables calculated from the privacy-preserving data aggregation mechanism proposed in Section~\ref{Section data-sharing}. Note that the formulations in this section are deterministic, i.e., without considering the parameter perturbations generated to ensure DP.

\subsubsection{Objective}
In this study, the objective function is to minimize the total delay of queued CVs. Note that we choose this simple objective function to illustrate the design of privacy-preserving traffic control algorithms. The proposed methodology can be readily generalized to consider other choices of optimization objectives, such as vehicle delay, queue length, etc., which can be easily estimated from the aggregated traffic variables (e.g., arrival rate). 

\begin{figure}[htbp]
	\centering
		\includegraphics[scale=.5]{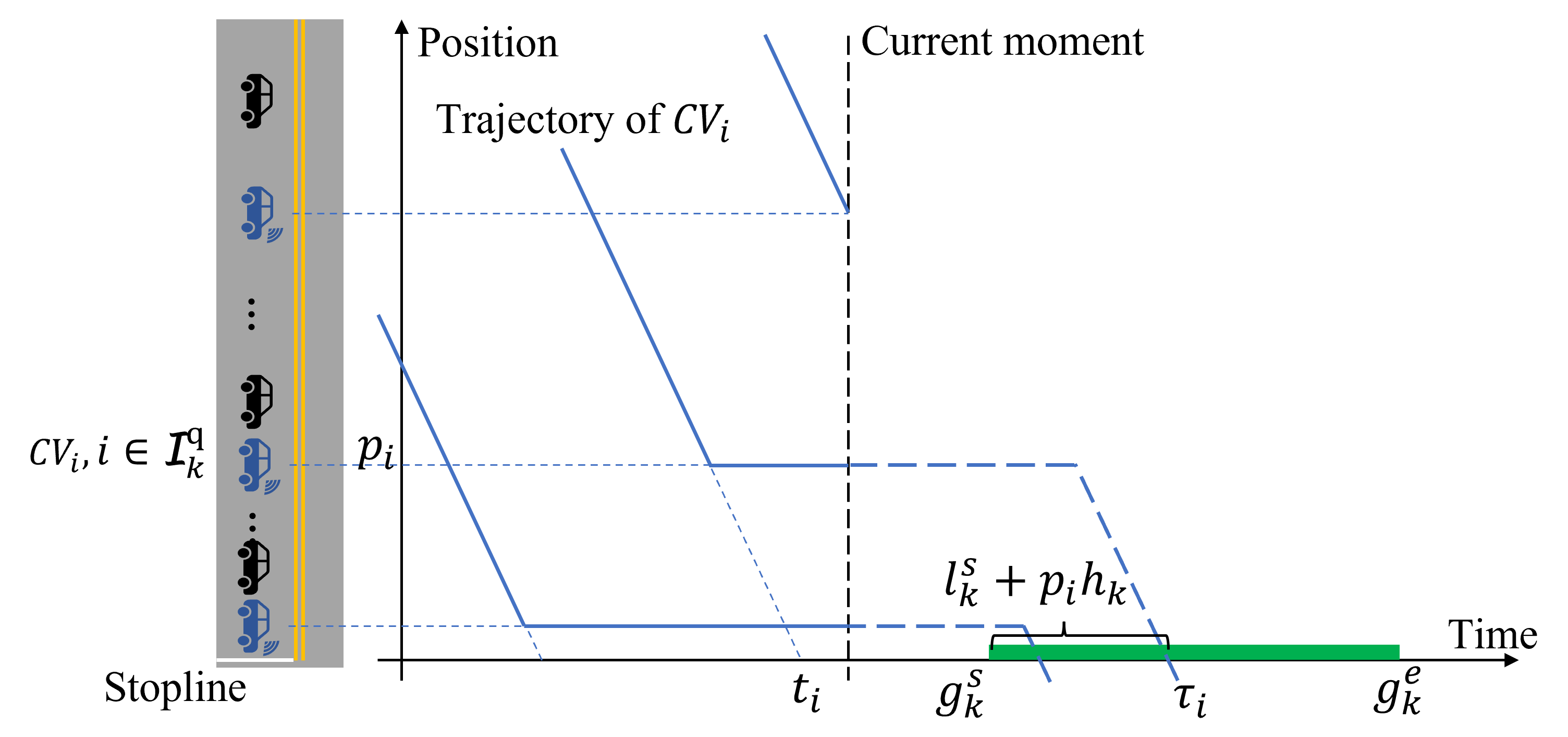}
	\caption{Calculation of the expected stopline through time of queued CVs}
	\label{FIG:5}
\end{figure}

Specifically, we use $g_k^s$ and $g_k^e$ to denote the green start time and green end time of stream $k$, respectively, which will be the decision variables of the optimization model. As shown in Fig. \ref{FIG:5}, the expected delay of a queued $CV_i$ in stream $k$ can be calculated as
\begin{eqnarray}\label{CVdelay}
    d_i=\tau_i-t_i,~i\in \mathcal{I}_{k}^{\rm{q}},
\end{eqnarray}
where $\mathcal{I}_{k}^{\rm{q}}$ indicates the set of queued CVs in stream $k$; $t_i$ is the expected arrival time at stopline of $CV_i$; $\tau_i$ is the expected stopline through time of $CV_i$, which can be written as 
\begin{eqnarray}\label{Throughtime}
    \tau_i=g_k^s+l_k^s+p_i h_k,~i\in \mathcal{I}_{k}^{\rm{q}},
\end{eqnarray}
where $l_k^s$ is the start-up lost time of stream $k$; $h_k$ is the average queue discharging headway of stream $k$ at the stopline; and $p_i$ is the current position of $CV_i$.

Thus, we have the total delay of queued CVs in stream k: 
\begin{eqnarray}\label{Totaldelay}
    \sum_{i\in \mathcal{I}_{k}^{\rm{q}}} d_i=\sum_{i\in \mathcal{I}_{k}^{\rm{q}}} (g_k^s+l_k^s+p_i h_k-t_i)=(g_k^s+l_k^s) N_{k,1}^{\rm{q}} + h_k \sum_{i\in \mathcal{I}_{k}^{\rm{q}}} p_i-\sum_{i\in \mathcal{I}_{k}^{\rm{q}}} t_i.
\end{eqnarray}

Similar to Eq. (\ref{lambda estimator 1}), we can let all CVs involve in the calculation of each stream parameter, then we have the following objective function,
\begin{eqnarray}\label{Obj}
\min \sum_{k\in \mathcal{K}} \sum_{i\in \mathcal{I}_{k}^{\rm{q}}} d_i=\sum_{k\in \mathcal{K}} (\eta_k (g_k^s+l_k^s)+h_k P_k-T_k).
\end{eqnarray}
\subsubsection{Constraints}
Note that we optimize the signal timing plan at the beginning of each phase group, thus we have two cases of the phase sequence: starting with streams $\{1,5\}$ and starting with streams $\{3,7\}$, shown in Fig. \ref{FIG:4}. Here we use the phase sequence starting with stream $\{1,5\}$ as an example to introduce the basic constraints that describe the logic of the NEMA ring-barrier structure:
\begin{align}
    & \sum_{k=1}^4 (g_k^e-g_k^s+y_k+a_k)=C, \label{Con1}\\
    & \sum_{k=5}^8 (g_k^e-g_k^s+y_k+a_k)=C, \label{Con2}\\
    & \sum_{k=1}^2 (g_k^e-g_k^s)=\sum_{k=5}^6 (g_k^e-g_k^s), \label{Con3}\\
    & g_k^e+y_k+a_k=g_{k+1}^s, \quad k=1,2,3,5,6,7  \label{Con4}\\
    & g_k^e-g_k^s \geq g_k^{\min}, \quad k=1,2,...,8  \label{Con5}\\
    & g_k^e-g_k^s \leq g_k^{\max}, \quad k=1,2,...,8  \label{Con6}\\
    & C_{\min} \leq C \leq C_{\max}, \label{Con7}
\end{align}
where $y_k$ is the yellow time allocated to stream $k$; $a_k$ is the red clearance time; $g_k^{\min}$ and $g_k^{\max}$ are the minimum and maximum green time, respectively; $C_{\max}$ and $C_{\min}$ are the upper and lower bounds of the cycle length, respectively. 

Eq. (\ref{Con1}) - Eq.(\ref{Con3}) ensure that the green time of each stream should follow the NEMA phase sequence; Eq. (\ref{Con4}) ensures that consecutive phases should have an inter-green time to ensure safety; Eq. (\ref{Con5}) - Eq. (\ref{Con7}) bound the green time and the cycle length. 

Note that if the phase sequence starts with streams $\{3,7\}$, Eq. (\ref{Con4}) should be
\begin{align}
\begin{cases}
 &g_k^e+y_k+a_k=g_{k+1}^s, \quad k=1,3,5,7\\
 &g_4^e+y_4+a_4=g_1^s, \\
 &g_8^e+y_8+a_8=g_5^s.
\end{cases}
\end{align}

As described in Section~\ref{Sec: JMLE}, we attempt to design signal timings such that each stream is under-saturated. This is to improve the operation of the signalized intersection (e.g., reduce the impact of lost time and eliminate the number of stops and residual queues) and to reduce the error of arrival rate estimation. Thus, we impose the following constraint: 
\begin{eqnarray}
    \frac{g_k^e-g_k^s+y_k-l_k^s-l_k^y}{h_k} \geq \lambda_k (g_k^s-r_k^{s,0}), \quad k \in \mathcal{K}, \label{Con8}
\end{eqnarray}
where the left-hand side of the inequality indicates the maximum number of vehicles that can be discharged during the effective green time represented as the difference between total green/yellow time $g_k^e-g_k^s+y_k$ and lost time (i.e., the sum of the start-up lost time $l_k^s$ and the yellow lost time $l_k^y$); the right-hand side of the inequality indicates the expected number of queued vehicles during the red time with $r_k^{s,0}$ representing the last red start time of stream $k$. Notice that the arrival rate $\lambda_k$ can be estimated by Eq. (\ref{lambda estimator 1}) or (\ref{lambda estimator 2}). Eq.(\ref{Con8}) requires that the queue length accumulated during the red time can be cleared during the effective green time. Nevertheless, in Section~\ref{subsubsec:model1}, we further relax the constraint to allow slightly oversaturated traffic conditions to ensure the feasibility of the optimization problem. 

In summary, we have established the following linear optimization model:  
\begin{align}
     \min\limits_{\theta} &\quad \sum_{k \in \mathcal{K}}\eta_k g_k^s, \label{OmittedObj} \\
   \text{s.t.} &\quad  \text{  Eq. (\ref{Con1}) - Eq. (\ref{Con8})}\nonumber,
\end{align}
where $\theta$ indicates the set of decision variables and $\theta = \{C, \{g_k^s,g_k^e \}_{k \in \mathcal{K}} \}$, and constants in the objective can be omitted.

\subsubsection{Model 1: linear programming with deterministic parameters} \label{subsubsec:model1}
We relax the requirement of under-saturated conditions by introducing decision variables $Q_k$ to convert Eq. (\ref{Con8}) into a soft constraint. Such treatment allows Eq. (\ref{Con8}) to be violated, which enables the following optimization problem P1 to consider over-saturated conditions.
\begin{align}
    P1:\quad  \min\limits_{\theta} \quad &\sum_{k \in \mathcal{K}} (\eta_k g_k^s+C_{max} Q_k), \label{P1Obj}\\
   \text{s.t.} \quad & \text{Eq. (\ref{Con1}) - (\ref{Con7})}, \nonumber \\ 
    & Q_k \geq \lambda_k (g_k^s-r_k^{s,0})-\frac{g_k^e-g_k^s+y_k-l_k^s-l_k^y}{h_k},\quad  k \in \mathcal{K}, \label{P1Con7}\\
    & Q_k \geq 0, \quad k \in \mathcal{K}. 
\end{align}
where $Q_k$ indicates the length of the residual queue (in terms of the number of vehicles) in stream $k$; $C_{max} Q_k$ in the objective indicates the maximum possible delay experienced by vehicles in the residual queue in stream $k$; $\lambda_k =F_{\lambda}(P_k, T_k)$ is the joint maximum likelihood for arrival rate, as indicated in  Eq. (\ref{lambda estimator 1}).

Since model P1 is a linear programming problem, we can solve it easily with solvers such as Gurobi and CPLEX.
The model P1 serves as the basic deterministic optimization problem, which directly takes the traffic variables as output from the privacy-preserving data aggregation mechanism. In scenarios where the privacy-preserving mechanism uses only SMPC, i.e., we can access accurate parameters of the aggregated information $\eta_k$, $P_k$, and $T_k$ from CVs, the resulting signal timings of the deterministic optimization problem will be the same as if no privacy-preserving mechanism is applied. In other words, the cost of preserving privacy is zero. Otherwise, in scenarios where the privacy-preserving mechanism uses both DP and SMPC, the output from the privacy-preserving data aggregation mechanism will be perturbed and becomes $\hat{\eta}_k$, $\hat{P}_k$, and $\hat{T}_k$, which may negatively affect the optimality of the resulting signal timings.

\subsection{Stochastic optimization model} \label{stochastic model}
This subsection aims to explicitly handle the noises brought by the LAP mechanism used to achieve DP. Specifically, at each decision step, we can only obtain perturbed traffic variables $\{\hat{\eta}_k,\hat{P}_k,\hat{T}_k\}_{k\in \mathcal{K}}$ without knowing the exact values of the system states $\{\eta_k, P_k, T_k\}_{k \in \mathcal{K}}$. In other words, we need to determine signal timing parameters with only prior knowledge of the distributions of these traffic variables. To this end, we formulate each decision step as a two-stage stochastic programming problem (see Section \ref{subsubsec:SP}), which will be approximated into a linear programming problem using a sample-path approach (see Section \ref{subsubsec:sampling}). 

\subsubsection{Two-stage stochastic programming} \label{subsubsec:SP}
We present the two-stage stochastic programming problem. The first-stage problem decides the signal timings subject to the deterministic constraints Eq.(\ref{Con1})-(\ref{Con7}) without knowing the realization of these random traffic variables. The second stage problem calculates the expected total costs considering the distribution of these traffic variables. Specifically, the two-stage stochastic programming problem is written as follows.  
\begin{align}
    \min\limits_{\theta} \quad & \mathbb E_{\omega}\left[ \hat{\mathscr{J}}(\theta, \omega) \right], \label{SPObj1}\\
    \text{s.t.} \quad &  \text{Eq. (\ref{Con1}) - (\ref{Con7}) }, \nonumber \\ 
    \text{where} \quad & \hat{\mathscr{J}}(\theta, \omega) = \min\limits_{\{Q_k(\omega)\}_{k \in \mathcal{K}}} \quad \sum_{k \in \mathcal{K}} \Big(\eta_k(\omega) g_k^s + C_{max} Q_k(\omega)\Big), \label{SPObj2} \\
    \text{s.t.} \quad & Q_k(\omega) \geq F_{\lambda}(P_k(\omega),T_k(\omega)) \cdot (g_k^s-r_k^{s,0})-\frac{g_k^e-g_k^s+y_k-l_k^s-l_k^y}{h_k}, \quad k \in \mathcal{K}, \label{SPCon9}\\
    &Q_k(\omega) \geq 0, \quad k \in \mathcal{K}.
\end{align}
where $\omega$ represents a random sample in the sample space $\Omega$, $\eta_k \sim Lap(\hat{\eta}_k,\Delta_k^{\eta} / \epsilon_k^{\eta})$, $P_k \sim Lap(\hat{P}_k,\Delta_k^P / \epsilon_k^P)$, and $ T_k \sim Lap(\hat{T}_k,\Delta_k^T / \epsilon_k^T)$ represent our \emph{a priori} knowledge of the exact values of the corresponding traffic variables. The first-stage decisions are made before the realization of random sample $\omega$ is observed, where the decision variables are represented as $\theta = \{C, \{g_k^s,g_k^e \}_{k \in \mathcal{K}} \}$. The second-stage decisions are made after the realization of random sample $\omega$ , where  the decision variables are represented as $\{ Q_k(\omega)\}_{k \in \mathcal{K}}$.  The second-stage objective function $\hat{\mathscr{J}}(\theta, \omega)$ indicates the sum of total delay and penalty of over-saturation, and the first-stage objective function $\mathbb E_{\omega}\left[ \hat{\mathscr{J}}(\theta, \omega) \right]$ indicates the expectation of $\hat{\mathscr{J}}(\theta, \omega)$ over the sample space $\Omega$. 

We make the following two remarks for the two-stage stochastic optimization problem. The first remark explains why \emph{a priori} knowledge of $P_k$, $T_k$, $\eta_k$ can be characterized as Laplace distributions. 
\begin{remark}[Prior knowledge of $P_k$, $T_k$, $\eta_k$]
Our prior knowledge of $P_k$, $T_k$, $\eta_k$ is two-fold. First, given the perturbed value $\hat{P}_k$, by Eq. (\ref{FinalP}), we have $P_k=\hat{P}_k-\xi_k^P$, where $\xi_k^P \sim Lap(0,\Delta_k^P / \epsilon_k^P)$, and hence we have $P_k \sim Lap(\hat{P}_k,\Delta_k^P / \epsilon_k^P)$. Similarly, given the perturbed value $\hat{T}_k$, we have $T_k \sim Lap(\hat{T}_k,\Delta_k^T / \epsilon_k^T)$, and  given the perturbed value $\hat{\eta}_k$, we have $\eta_k \sim Lap(\hat{\eta}_k,\Delta_k^{\eta} / \epsilon_k^{\eta})$. Second, since all traffic variables are physically bounded, we have $P_k, \eta_k, T_k\geq 0$, and the estimation of arrival rate $F_{\lambda}(P_k,T_k)$ should be within a certain predefined range. 
\end{remark}

The second remark can be used to simplify the objective functions. 
\begin{remark}[Simplification of objective functions]
    We further note that Eq.(\ref{SPObj1}) can be rewritten as $\mathbb E_{\omega} \left[ \hat{\mathscr{J}}(\theta, \omega) \right] = \mathbb{E}_{\omega}\left[\sum_{k \in \mathcal{K}} \Big(\eta_k(\omega) g_k^s + C_{max} Q_k(\omega)\Big)\right]=\sum_{k \in \mathcal{K}} \hat{\eta}(\omega)g_k^s + C_{max} \mathbb E_{\omega} \left[\sum_{k \in \mathcal{K}}Q_k(\omega) \right]$. Hence, the first-stage objective function can be re-written as 
    \begin{align}
        \min\limits_{\theta} \quad & \mathbb \sum_{k \in \mathcal{K}} \hat{\eta}
_kg_k^s + C_{max}  \mathbb{E}_{\omega}\left[ \hat{Q}(\theta, \omega) \right],
    \end{align}
    and the second-stage objective function can be re-written as 
    \begin{align}
        \hat{Q}(\theta, \omega) = \min\limits_{\{Q_k(\omega)\}_{k \in \mathcal{K}}} \quad \sum_{k \in \mathcal{K}} Q_k(\omega).
    \end{align}
\end{remark}

\subsubsection{Model 2: sampling-based linear programming with stochastic parameters} \label{subsubsec:sampling}
To solve the proposed two-stage stochastic programming problem, we approximate it to a linear programming problem using a sample-path approach. Specifically, we can generate a sufficient number of scenarios by Monte Carlo sampling, i.e., $\{P_k^m,T_k^m\}_{m \in \mathcal{M}}$, to characterize the uncertainties of parameters, where $\mathcal{M}$ represents the set of sampled scenarios. The resulting sampled scenarios should be physically meaningful, i.e., $P_k^m, T_k^m\geq 0$, and the estimation of arrival rate $F_{\lambda}(P_k^m,T_k^m)$ should be within a pre-defined range. To achieve this, we remove the sampled scenario $m$ and resample if these conditions are violated. 
Note that due to the simplification of objective functions in Remark 2, the optimization problem does not include $\eta_k$. Then the approximated linear programming problem can be written as 
\begin{align}
   P2:\min\limits_{\theta, \{ Q_k^m\}_{k \in \mathcal{K}, m \in \mathcal{M}}} \quad & \sum_{k \in \mathcal{K}} \hat{\eta}_k g_k^s+\frac{C_{max}}{| \mathcal{M}|} \sum_{m \in  \mathcal{M}} \sum_{k \in \mathcal{K}} Q_k^m, \label{P2Obj}\\
    \text{s.t.} \quad &  \text{Eq. (\ref{Con1}) - (\ref{Con7})}, \nonumber \\ 
    & Q_k^m \geq F_{\lambda}(P_k^m,T_k^m) \cdot (g_k^s-r_k^{s,0})-\frac{g_k^e-g_k^s+y_k-l_k^s-l_k^y}{h_k},  k \in \mathcal{K}, m \in  \mathcal{M} \label{P2Con9} \\  
    & Q_k^m \geq 0, \quad k \in \mathcal{K}, \quad m \in  \mathcal{M}. \label{P2Con10}
\end{align}
The linear programming problem P2 can be solved efficiently with solvers such as Gurobi and CPLEX.  

\section{Determination of $\epsilon$ and Sensitivity of DP}  \label{Section parameter}
\subsection{Choice of $\epsilon$} \label{choice of Pdire}
In DP, the parameter $\epsilon$ determines the magnitude of the added perturbation and hence controls the crucial tradeoff between the accuracy of the query results and the strength of the privacy guarantee. Existing studies have shown that the impact of $\epsilon$ on the query results varies across query problems \cite{Lee2011}, and thus the choice of $\epsilon$ should be problem-specific. Therefore, in this section, we aim to explore how to choose $\epsilon$ within our privacy-preserving traffic signal control framework. 

Based on the study by \cite{Lee2011} and \cite{Mehner2021}, given a database consisting of $N$ individual data subjects (i.e., CVs in our case) and the maximum allowed probability $P_{risk}$ for a particular individual being identified in the database after a query, the parameter $\epsilon$ should satisfy
\begin{align}
    \epsilon \leq \ln \frac{P_{risk}(N-1)}{1-P_{risk}}, \label{Prisk}
\end{align}

In particular, we require $P_{risk} > 1/N$ such that $\epsilon>0$. When $P_{risk}=1/N$ (i.e., the maximum allowed probability of identifying a particular individual in the database is equivalent to that of a random guess), the query results should degrade to pure random noise that consists of no information and thus provides no data utility. 
 
Note that in the proposed data aggregation mechanism, all CVs within a zone of interest around the studied intersection will participate in the calculation of query results. Compared to the risk of the existence of a CV being identified at the studied intersection, we are more concerned about the risk that a CV is identified on a specific link and direction, as the link/direction information can reveal the route that the CV takes. Therefore, the maximum allowed probability of a CV being identified in a specific direction (denoted by $P_{dire}$) is one-eighth of $P_{risk}$  since a four-link intersection has 8 possible directions (i.e., 4 links each with 2 directions). Thus, we have 
\begin{align}
    \epsilon \leq \ln \frac{8P_{dire}(N-1)}{1-8P_{dire}}, \label{Pdire}
\end{align}

Eq. (\ref{Pdire}) indicates that under the same level of maximum allowed risk, a larger $N$ gives $\epsilon$ a larger range of values. This also suggests that the proposed data aggregation mechanism that involves all CVs within the zone of interest in the calculation of each stream parameter generates less perturbation on the aggregation results than the usual treatment that uses only CVs in the corresponding stream.

In our cases, the total number of CVs within the zone of interest $N$ is unknown before we aggregate CV data, thus the historical average of $N$ (also perturbed by the proposed mechanism) is adopted to determine the upper bound of $\epsilon$.

\subsection{Sensitivity of vast majority} \label{Sensivity of vast majority}
In DP, the magnitude of the added perturbation strongly depends on the sensitivity of the query function given $\epsilon$. 
It is evident that the global sensitivity of $\eta_k$ is $\Delta^{\eta}_k=1$ since adding/removing a CV will change the counting variable $\eta_k$ by at most 1. Similarly, the global sensitivity of $P_k$ and $T_k$ should be the maximum values of $p_{k,i}$ and $t_{k,i}$, respectively, which can be written as 
\begin{align}
    \Delta^{P}_k = \max\left\{ p_{k,i} | i \in \mathcal{I} \right\} < \frac{L_{link}}{L_0}, \quad \Delta^{T}_k = \max\left\{ t_{k,i} | i \in \mathcal{I} \right\} <C_{\max}, \label{UB}
\end{align}
where $L_{link}$ is the maximum length of incoming links. Recall that $L_0$ represents the spacing of one vehicle, and $C_{\max}$ represents the maximum cycle length. 
 
 However, the bounds of sensitivity $C_{\max}$ and $L_{link}/L_0$ can be too loose for practical applications, potentially resulting in large perturbations affecting the control performance. 
 To address this issue, we relax the DP requirement and adopt the idea of DP for the vast majority \cite{Kartal2019}, as described in Definition~\ref{dfn:DPVM}.
\begin{definition} [$(\epsilon,\alpha^r)$-Differential privacy] \label{dfn:DPVM}
    Given a dataset D that is divided into two non-overlapping sets $D_A$ and $D_B$ with $|D_A| \geq (1-\alpha^r)|D|$, a randomized perturbation mechanism $\mathcal{A}: \mathcal{D} \rightarrow \mathcal{X}$ satisfies $(\epsilon,\alpha^r)$-differential privacy, if for any neighboring databases $D_{A,1} \subseteq D_A$ and $D_{A,2} \subseteq D_A$ that differ on one record (e.g., adding or removing one record), and any possible output set $\mathcal{E} \subset \mathcal{X}$, we have 
\begin{align}
    Pr[\mathcal{A}(D_{A,1})\in \mathcal{E}]\leq e^\epsilon Pr[\mathcal{A}(D_{A,2})\in \mathcal{E}]. \label{DPVM}
\end{align}
    where $|\cdot|$ indicates the number of records in the dataset. 
\end{definition}

The core idea of $(\epsilon,\alpha^r)$-DP is to obtain better query results by reducing the global sensitivity of the query function at the cost of sacrificing the privacy of a small portion (i.e., $\alpha^r$) of individuals. 
$(\epsilon,\alpha^r)$-DP ensures that at least $1-\alpha^r$ proportion of individuals in the dataset (i.e., $D_A$) satisfies $\epsilon$-DP with mechanism $\mathcal{A}$. For the remaining smaller proportion of individuals (i.e., $D_B$), $\epsilon$-DP is not satisfied, but their privacy can still be protected at a lower level (i.e., with a larger DP parameter $\epsilon$) by the added perturbations.

Normally, $\alpha$ is determined based on the distribution of the dataset $D$, and $r$ is a user-specified parameter no smaller than 1. However, in our cases the real private data of CVs is unknown, thus the $\alpha$ cannot be calibrated based on the dataset. Therefore, in this study, we implement the DP for the vast majority by empirically adopting a sensitivity $\Delta_{f,vast}$.  
It is worthwhile noting that, the privacy sacrifice of the small proportion of CVs is acceptable in our method because we still have SMPC to protect their real private data.

For $T_k$, we use a red time $R_k$ related parameter as the sensitivity of the LAP mechanism in DP, since most of the queued vehicles arrived at the intersection during the red time. For $P_k$, an empirically determined queue length value is adopted. In summary, we have
\begin{align}
    \Delta_k^\eta = 1, \quad \Delta_k^P = Q_e, \quad \Delta_k^T = \varphi R_k, \label{Sensitivity}
\end{align}
where $Q_e$ is an empirically determined value and $\varphi$ is a coefficient greater than 0 to control the sensitivity of $T_k$, both of which will be discussed in Section~\ref{Section evaluation}.

\section{Evaluation}  \label{Section evaluation}
\begin{figure}[htbp]
	\centering
		\includegraphics[scale=.55]{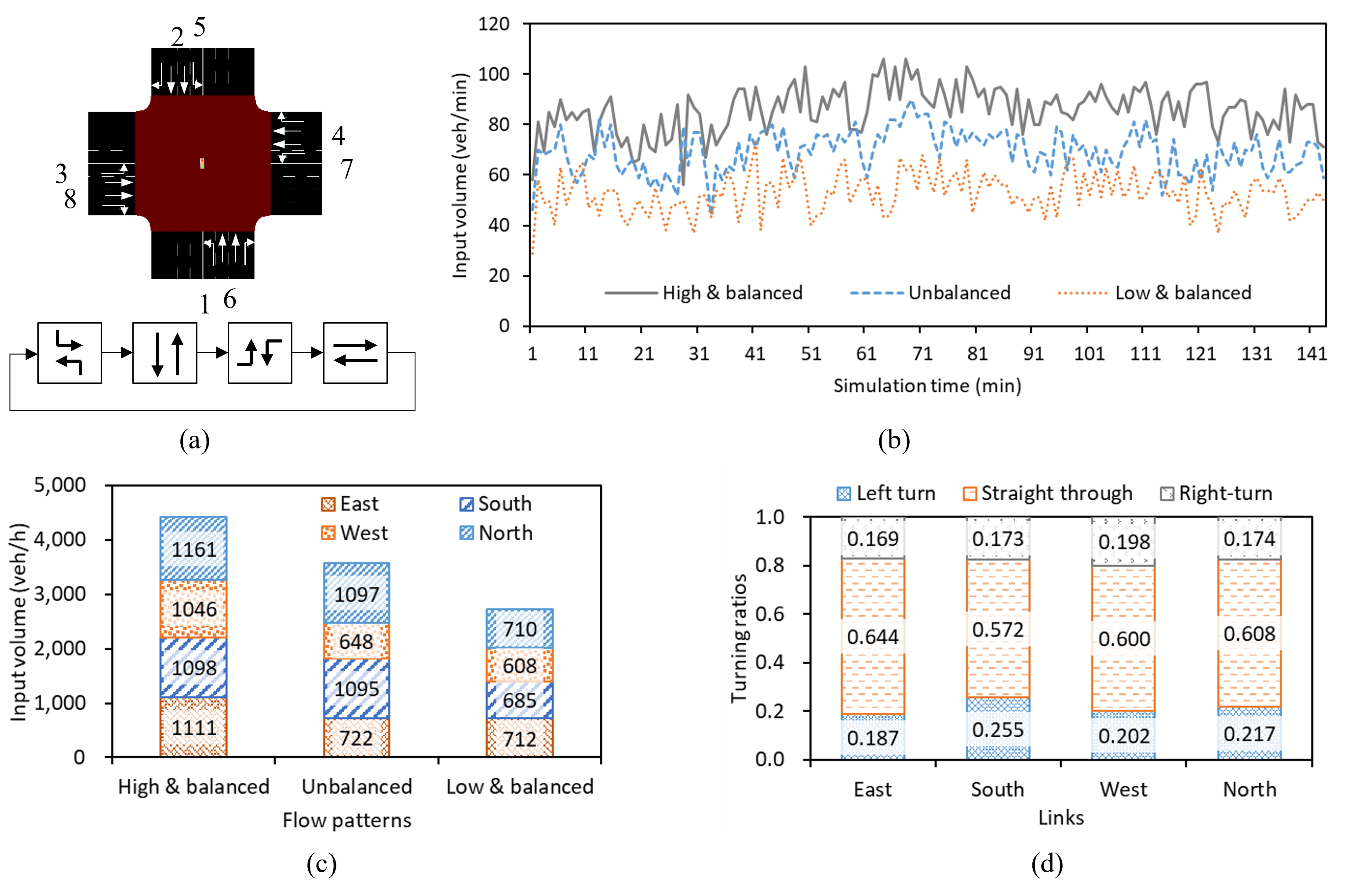}
	\caption{Basic settings of the simulation: (a) intersection layout and phase sequence; (b) intersection volume per minute of different flow patterns; (c) total volume of different flow patterns; (d) turning ratios of each link.}
	\label{FIG:6}
\end{figure}

The proposed methods are evaluated at a signalized intersection simulated in SUMO. The intersection layout and phase sequence are shown in Fig. \ref{FIG:6}\,(a). All links have two straight-through lanes, one left-turning lane, and one uncontrolled right-turning lane. The phase sequence consists of four phases with leading left turns. The simulation time is 10,000\,s, where the 2-hour data from 1,300\,s to 8,500\,s is used for evaluation. 

We consider three types of flow patterns: (1) high and balanced demand, (2) low and balanced demand, and (3) unbalanced demand. For each flow pattern, vehicle arrival at each link is characterized as a Poisson process with time-dependent link demand presented in Fig.\ref{FIG:6}\,(c). 
Each arrived vehicle will decide its route (i.e., through-moving, left-turning, and right-turning) based on the turning probabilities summarized in Fig.\ref{FIG:6}\,(d). Note that such settings can generate rather stochastic demand, as the total demand at the intersection fluctuates around 3000 veh/h over the entire simulation period for three types of flow patterns (see Fig.\ref{FIG:6}\,(b)). 

The proposed methods are tested in scenarios with various penetration rates, ranging from 0.2 to 0.9 with a step size of 0.1. The input volume and turning ratios are the same for all scenarios. CVs are sampled from the entire vehicle population following binomial distribution (i.e., the probability of each vehicle being set as connected equals the penetration rate). For each penetration rate, the experiments will be repeated 10 times with different random seeds for sampling CVs.

\subsection{Benchmarks}
The following traffic signal control methods are tested for comparison:\vspace{-0.3em}
\begin{itemize}[leftmargin=2em]
    \item Actuated: A gap-based fully-actuated traffic control method embedded in SUMO, which switches phases after detecting a sufficient time gap between successive vehicles. In our simulation, the maximum gap is adopted as 3 s. Note that this method assumes the existence of loop detectors, while the other CV-based benchmarks assume no loop detectors. Moreover, this method does not provide a DP guarantee for the privacy of vehicles passing the loop detectors. \vspace{-0.3em}
    \item LP: The basic linear programming model with accurate deterministic parameters (P1). In other words, this method assumes that the input data to P1 does not have any noise. It can also be seen as applying only SMPC to protect the private data of CVs without integrating with DP. \vspace{-0.3em}
    \item Privacy-LP: The basic deterministic linear programming model (P1) that directly takes the perturbed parameters as input and treats them as deterministic. This method can be seen as scenarios where both SMPC and DP are applied to the private data of CVs. The comparison between Privacy-LP and LP evaluates the sensitivity of LP to noises generated to ensure DP.\vspace{-0.3em}
    \item Privacy-TSP: The stochastic linear programming model (P2) that treats the input as stochastic parameters. Both SMPC and DP are applied to protect the private data of CVs, and the noises generated to ensure DP are explicitly handled. The comparison between Privacy-TSP and Privacy-LP shows the benefits of stochastic programming. \vspace{-0.3em}
\end{itemize}

The minimum green time is 10 s, and the maximum green time is 60 s for all methods. The red clearance time is not considered, and the yellow time is 3 s. The nominal values of related parameters of the proposed methods are first given for performance comparison: privacy-related parameters $P_{dire}=0.05$ (appears in Eq. (\ref{Pdire})), $\varphi=1$ and $Q_e=8$ (appears in Eq. (\ref{Sensitivity})); stochastic programming-related parameter $|\mathcal{M}|=400$ (number of scenarios, appears in Eq. (\ref{P2Obj})). The nominal flow pattern is high and balanced demand.

\subsection{Performance of different traffic signal control methods} \label{overall performance}
\subsubsection{Different penetration rates} \label{sec: penetration rate}
\begin{figure}[htbp]
	\centering
		\includegraphics[scale=.6]{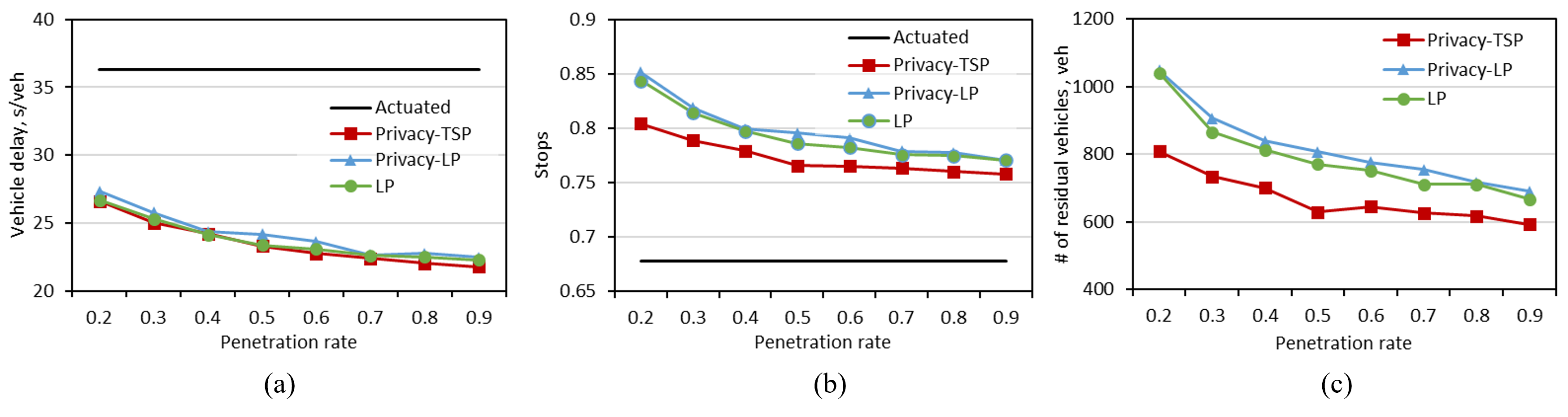}
	\caption{The performance of four test methods under different penetration rates: (a) vehicle delay; (b) number of stops; (c) number of residual vehicles.}
	\label{FIG:7}
\end{figure}

In this section, we test the performance of the proposed methods under various penetration rates. 
Fig. \ref{FIG:7} illustrates the performance of the four methods on performance indicators such as average vehicle delay, number of stops, and number of residual vehicles (i.e., vehicles carried over to the next cycle). 

\emph{Value of CV information}. Compared to actuated control, the three CV-based methods significantly reduce average vehicle delay in scenarios with all penetration rates, which shows the value of using CV information. Nevertheless, we can observe an increase in the number of stops and the number of residual vehicles.
This is expected because actuated control typically switches the green phase after the queue in the stream dissipates, resulting in little to no residual queues in that stream, which could be detrimental to vehicle delays in the other streams.
Moreover, as the penetration rate increases, the vehicle delay, the number of stops, and the number of residual vehicles decrease for all three CV-based methods, i.e., LP, Privacy-LP, and Privacy-TSP. This is expected because the increase in penetration rates can improve the information acquired by the signal controller. We also notice that the marginal benefits of having more CVs decrease as the penetration rate increases, which shows that the value of CV information is more significant in the early adoption period. 

\emph{Control impact of privacy protection}. The impact of privacy protection is evaluated by comparing Privacy-LP and LP. Recall that Privacy-LP 
 solves P1 with input data from the data aggregation mechanism combining DP and SMPC, whereas LP represents the scenario with accurate input data (i.e., either with SMPC only or no privacy protection at all). Evidently, Privacy-LP is found to perform similarly to LP, as the difference between these two methods in all performance indicators is marginal. This implies that Privacy-LP can ensure the privacy of CVs with a marginal impact on control performance. 

\emph{Value of employing stochastic programming}. We demonstrate the value of employing stochastic programming to explicitly handle DP noises by comparing Privacy-TSP and Privacy-LP. We can see that Privacy-TSP can significantly reduce the number of stops and the number of residual vehicles while ensuring a slightly better average vehicle delay. This implies that Privacy-TSP is more effective in alleviating oversaturated traffic conditions by explicitly handling the noise generated to ensure DP through stochastic programming.

\subsubsection{Different flow patterns} \label{sec: flow pattern}
\begin{figure}[htbp]
	\centering
		\includegraphics[scale=.7]{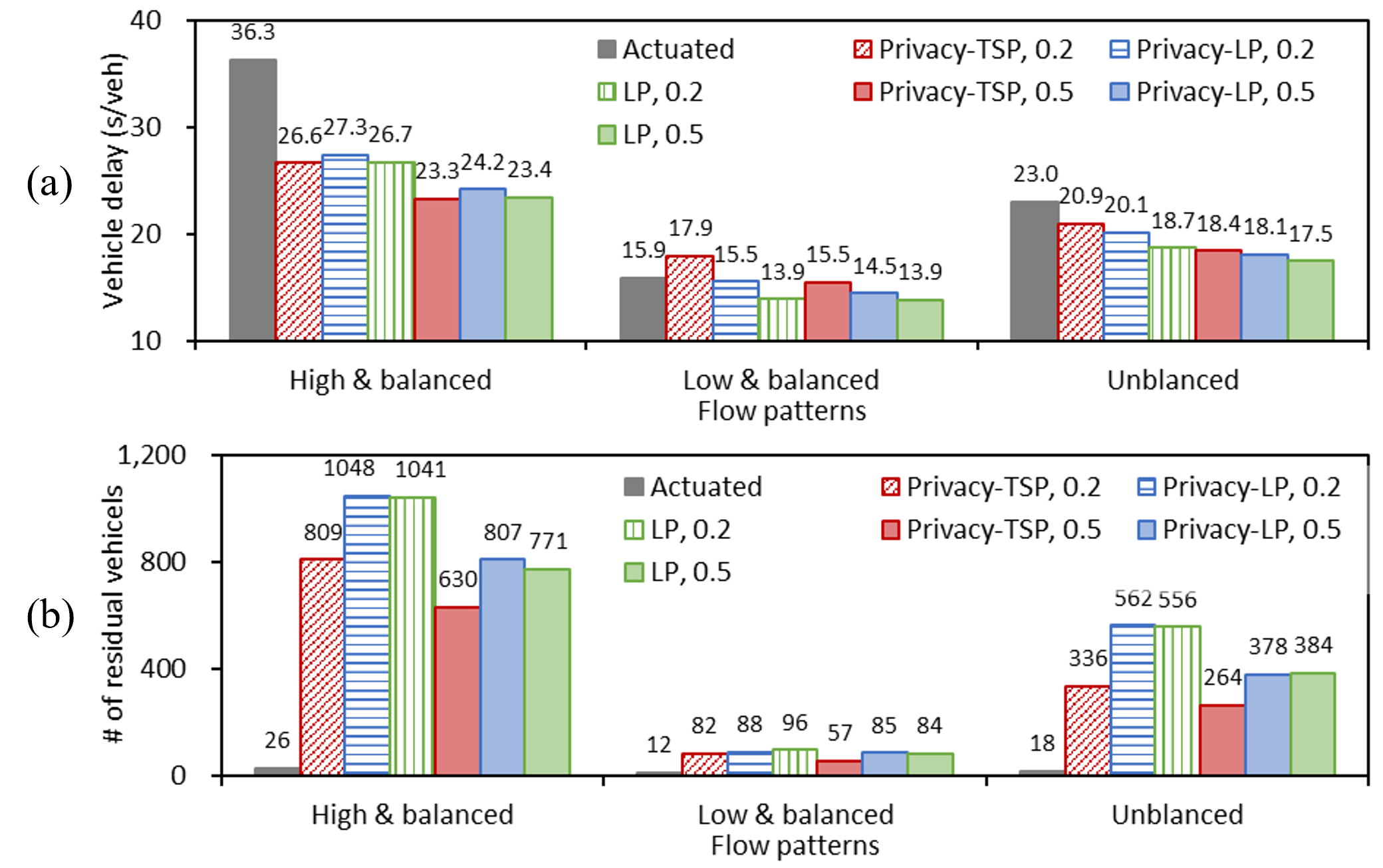}
	\caption{The performance of four test methods under different flow patterns: (a) vehicle delay; (b) number of residual vehicles.}
	\label{FIG:flow patterns}
\end{figure}

We evaluate the performance of the proposed methods under different flow patterns. The penetration rates of 0.2 and 0.5 are adopted to represent the adoption of CVs in the near and medium-term future, and the results are shown in Fig. \ref{FIG:flow patterns}.

All three proposed methods are found to perform better under higher traffic demand levels, as the improvement of vehicle delays compared to actuated control is more significant. 
This is expected because the noises added by DP have less impact as the number of vehicles increases. In other words, it is easier to ``hide'' a CV in a larger population. 
Moreover, we note that Privacy-TSP performs worse than Privacy-LP in scenarios with the low-and-balanced flow pattern and unbalanced flow pattern, even worse than actuated control in scenarios with the low-and-balanced flow pattern and a penetration rate of 0.2. This is because fewer CVs are involved in data sharing at lower traffic demand levels and penetration rates, and the relative error introduced by DP is larger as the aggregated data is smaller, which leads to a larger bias in the distribution of arrival rates and downgrades performance of stochastic programming. In particular, the proposed three methods still significantly reduce vehicle delay in unbalanced traffic flow patterns compared to the actuated control, which demonstrates the adaptability of the proposed methods.

\subsection{Privacy-utility tradeoff}
We investigate the privacy-utility tradeoff by evaluating the relationship between privacy and control performance. The control performance of our proposed method is significantly influenced by the magnitude of the Laplace perturbations, which is controlled by the privacy budget of DP $\epsilon$ and the sensitivity of the vast majority for parameters $\hat{T}_k$ and $\hat{P}_k$. By Eq. (\ref{Pdire}), the privacy budget $\epsilon$ is calculated using a pre-determined allowable level of privacy risk, characterized as the probability of a CV being identified in a specific direction $P_{dire}$. The sensitivity of the vast majority for parameters $\hat{T}_k$ and $\hat{P}_k$, on the other hand, can be obtained by coefficients $\varphi$ and $Q_e$, which determine the percentage of users for whom $\epsilon$-DP can be satisfied. Hence, We will perform sensitivity analysis for privacy risk $P_{dire}$ (Section~\ref{subsubsec:pdire}) and sensitivity parameters $\{\varphi, Q_e\}$ (Section~\ref{subsubsec:sensitivity}). The penetration rate is set to 0.5, while other parameters are set to the nominal values except parameters to be tested.

\subsubsection{Risk for identifying CV direction, $P_{dire}$} \label{subsubsec:pdire}
We perform sensitivity analysis to $P_{dire}$.  To this end, we first identify representative values of $P_{dire}$ to be tested. As mentioned in Section \ref{choice of Pdire}, the probability $P_{dire}= P_{risk} / 8 \leq 0.125$. Moreover, the choice of $P_{dire}$ should ensure the resulting $\epsilon>0$ (i.e., by Eq. (\ref{Pdire})), which gives $P_{dire}>0.003$ since the number of CVs, $N$,  participating in the optimization at each decision step in scenarios with a penetration rate of 0.5 is around 40-70. Therefore, we choose three representative values for $P_{dire}$: 0.01, 0.05, and 0.1. 

\begin{figure}[htbp]
	\centering
		\includegraphics[scale=0.6]{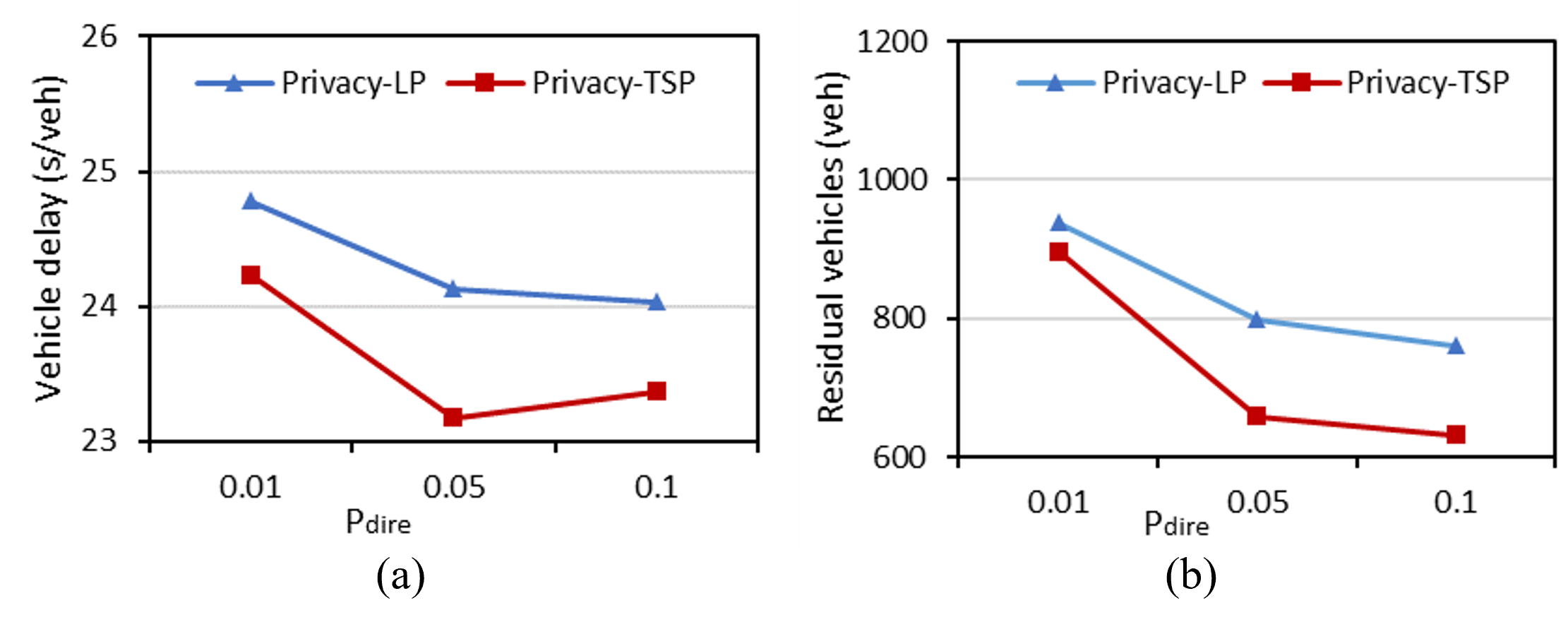}
	\caption{The performance of the proposed methods with different $P_{dire}$: (a) vehicle delay; (b) number of residual vehicles.}
	\label{FIG:Prisk}
\end{figure}

The control performance of Privacy-TSP and Privacy-LP with different $P_{dire}$ is shown in Fig. \ref{FIG:Prisk}. 
As $P_{dire}$ increases, both Privacy-TSP and Privacy-LP exhibit a decreasing trend in vehicle delay and the number of residual vehicles. However, the difference between the results of $P_{dire}=0.05$ and $P_{dire}=0.1$ is not significant, indicating that a further increase in $P_{dire}$ may not necessarily result in further improvements for Privacy-TSP. This can be explained by the characteristics of the introduced Laplace noises, for which we illustrate with a simple example of adding noise to $P_k$. We assume that $N=50$ and $Q_e=8$. By Eq. (\ref{Pdire}) and Proposition \ref{prp:LAP}, we know that $P_{dire}$ determines the upper bound of $\epsilon$, which serves as the denominator for the scale parameter $b$ of Laplace noise. Then we can calculate that $b$ is equal to 5.51, 2.30, and 1.51 when $P_{dire}$ is equal to 0.01, 0.05, and 0.1, respectively. It can be seen that when $P_{dire}$ increases from 0.01 to 0.05, the scale $b$ of Laplace noise decreases by 3.21, resulting in a significant improvement of control performance of the proposed methods; however, when $P_{dire}$ increases from 0.05 to 0.1, $b$ decreases by only 0.79, resulting in similar control performance of the proposed methods.


Moreover, we find that Privacy-TSP consistently outperforms LP in terms of vehicle delay and the number of residual vehicles under different $P_{dire}$ values. This result is consistent with the findings from Section \ref{overall performance} for the high and balanced flow pattern scenario, which suggest that stochastic programming is particularly effective in scenarios with higher traffic demands, where the aggregated data is larger.

\subsubsection{Sensitivity for the vast majority, $\varphi$ and $Q_e$} \label{subsubsec:sensitivity}
Recall that the LAP mechanism for DP depends on the notion of global sensitivity of the query function (see Proposition~\ref{prp:LAP}). To improve control performance, we relax this notion from the global sensitivity of all participants (see Definition \ref{dfn:sensitivity}) to the sensitivity of the vast majority (see Definition \ref{dfn:DPVM}). We next numerically investigate the impact of such a treatment on the privacy-utility tradeoff. Specifically, we evaluate the control performance of the proposed methods with 49 combinations of $Q_e$ and $\varphi$, i.e., $\{Q_e,\varphi \, | \, Q_e=2,4,...,14; \varphi = 0.2,0.4,...,1.4 \}$, where each combination is tested 3 times with different random seeds. 

Given a combination of $Q_e$ and $\varphi$, the CVs participating in data aggregation can be classified into three vehicle groups according to the protection level of $P_{k,i}$ and $T_{k,i}$ they shared:
\begin{itemize} 
    \item Type 1: CVs with $\epsilon-$DP guarantees in both parameters of $P_{k,i}$ and $T_{k,i}$; \vspace{-0.3em}
    \item Type 2: CVs with $\epsilon-$DP guarantee in one of the parameters $P_{k,i}$ and $T_{k,i}$ and lower-level  privacy protection in the other parameter. \vspace{-0.3em}
    \item Type 3: CVs with lower-level privacy protection in both $P_{k,i}$ and $T_{k,i}$. \vspace{-0.3em} 
\end{itemize}

\begin{figure}[htbp]
	\centering
		\includegraphics[scale=.5]{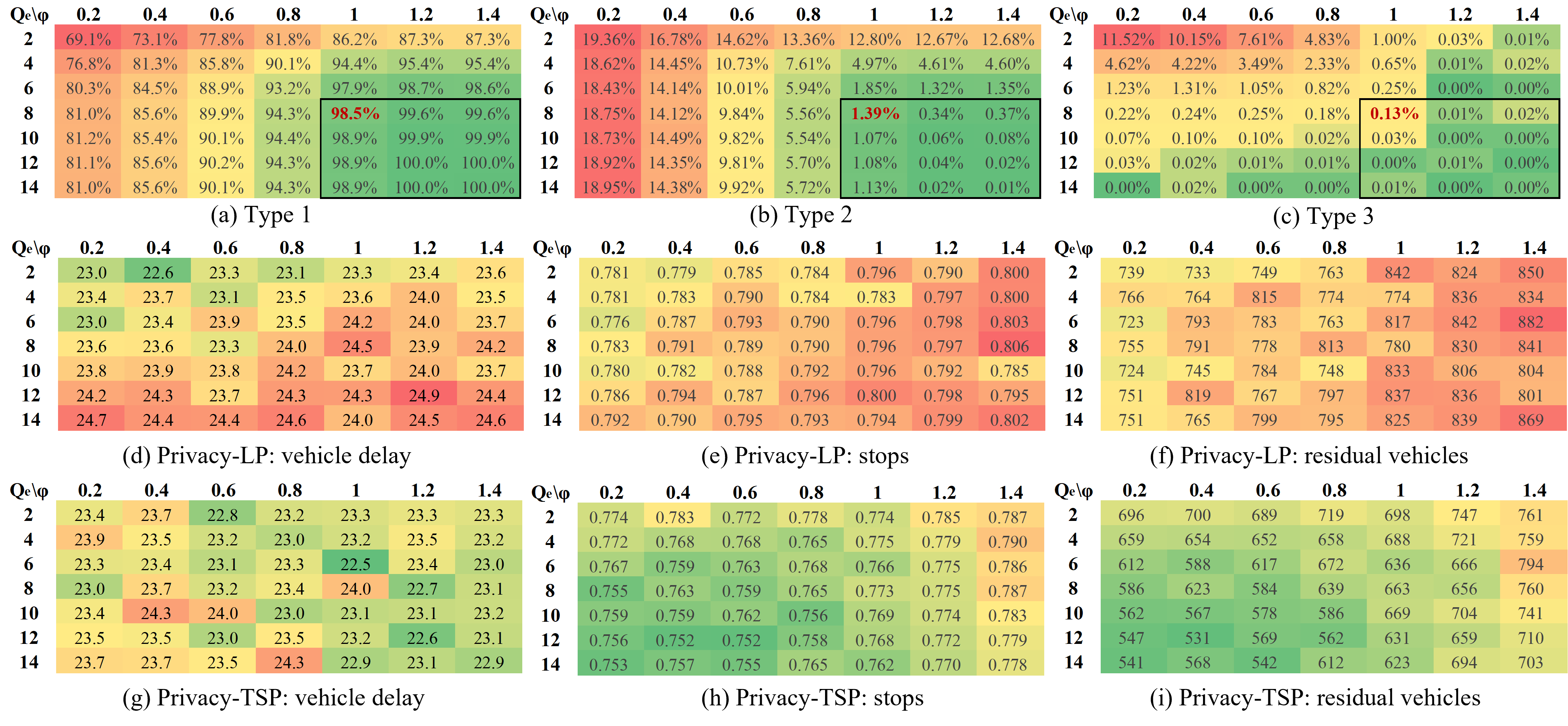}
	\caption{The performance of the proposed methods with different $\varphi$ and $Q_e$: (a)-(c) the proportions of three types of CVs; (d)-(f) the performance of Privacy-LP; (g)-(i) the performance of Privacy-TSP.}
	\label{FIG:9}
\end{figure}

Fig. \ref{FIG:9}(a)-(c) show the proportion of three types of CVs resulting from different combinations of $Q_e$ and $\varphi$. We can obtain the preferable ranges of these parameters, marked in the box with black borders. As we can see, larger values of $Q_e$ and $\varphi$ will result in more Type-1 CVs, i.e. CVs with $\epsilon-$DP guarantees. Specifically,  for $\varphi \geq 1$ and $Q_e \geq 8$, the proportion of Type-1 CVs is more than 98.5\%, indicating that $\epsilon-$DP of over 98.5\% of CVs can be guaranteed. It is important to note that the privacy of Type-2 CVs and Type-3 CVs is also protected by SMPC and DP with a lower-level guarantee, and hence it can still be challenging to infer private data from them.  

Fig. \ref{FIG:9}(d)-(f) show the performance of Privacy-LP under different combinations of $Q_e$ and $\varphi$, which reflect the impact of privacy-preserving mechanisms on the linear model P1. As is shown, with the increases in $Q_e$ and $\varphi$, both the vehicle delays, the number of stops, and the number of residual vehicles increase overall. The underlying reason is that the larger $Q_e$ and $\varphi$ bring more errors to the arrival rate estimation, which in turn reduces the utility of the linear model P1. 

On the other hand, the performance of Privacy-TSP under different combinations of $Q_e$ and $\varphi$ shows different trends from Privacy-LP, as presented by Fig. \ref{FIG:9}(g)-(i). With the increases in $Q_e$ and $\varphi$, the vehicle delay is decreasing overall; when vehicle delay is lower, the number of stops and the number of residual vehicles are usually larger. 
The former is because Privacy-TSP takes into account the distribution, i.e., prior knowledge, of the parameters through stochastic programming. When more errors are introduced by larger $Q_e$ and $\varphi$, the stochastic programming model considers more scenarios of arrival rates and thus has a greater probability of covering the true arrival rates and therefore achieves better performance, i.e., less vehicle delay. 
The latter is because our optimization objectives weigh the in-cycle delays of queued CVs and the expectation of the total number of residual vehicles, while they optimize the decision variables in different directions, i.e., shorter green lights lead to smaller in-cycle vehicle delays, yet larger residual vehicles. Thus, Privacy-TSP shows opposite trends for delays and stops/residual vehicles.

In summary, the deterministic method Privacy-LP usually requires a compromise between privacy and control performance, while the stochastic method Privacy-TSP does not.

\subsection{Sensitivity analysis for number of scenarios, $|\mathcal{M}|$}
We evaluate the impact of the number of scenarios $|\mathcal{M}|$ in the stochastic programming model on the control performance of Privacy-TSP in order to justify our choice of $|\mathcal{M}|$. 
\begin{figure}[htbp]
	\centering
        \includegraphics[scale=.6]{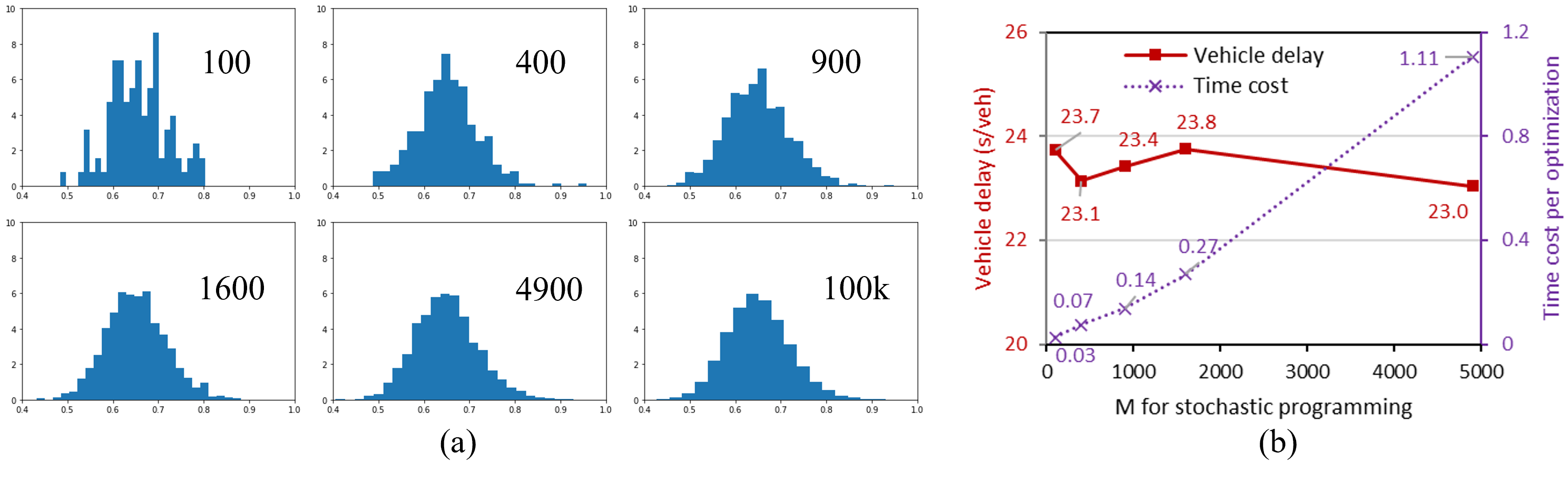}
	\caption{The performance of Privacy-TSP under different $\mathcal{M}$: (a) Distribution of $\lambda_k$; (b) Vehicle delay and time cost.}
	\label{FIG:11}
\end{figure}

The number of scenarios $|\mathcal{M}|$ significantly influences how accurately the sampled values of $\lambda_k$ can represent its distribution, which in turn affects the accuracy of the optimization problem P2 in approximating the two-stage stochastic programming problem established in Section \ref{subsubsec:SP}. As shown in Fig \ref{FIG:11}(a), with over 400 sampled scenarios of $P_k$ and $T_k$, the obtained empirical distribution of $\lambda_k$ is already very close to the empirical distribution with 100k sampled scenarios (treated as true distribution). The results in Fig. \ref{FIG:11}(b) indicate that Privacy-TSP showed little difference ($<0.8$ s/veh) in vehicle delay, while the time cost for solving P2 increases almost linearly with $|\mathcal{M}|$. 

Therefore, we adopted $|\mathcal{M}|=400$ in our cases, which meets the minimum number of scenarios required for stochastic optimization while solving P2 efficiently for real-time traffic signal control.

\subsection{Robustness of joint arrival rate estimator} \label{Sec: Performance estimator}
We test the robustness of the joint arrival rate estimator to the noises brought by the LAP mechanism. Note that the accuracy of the joint arrival rate estimator with different penetration rates, volume levels, and arrival patterns has been tested by \cite{Tan2022-est}.  

\begin{figure}[htbp]
	\centering
		\includegraphics[scale=.6]{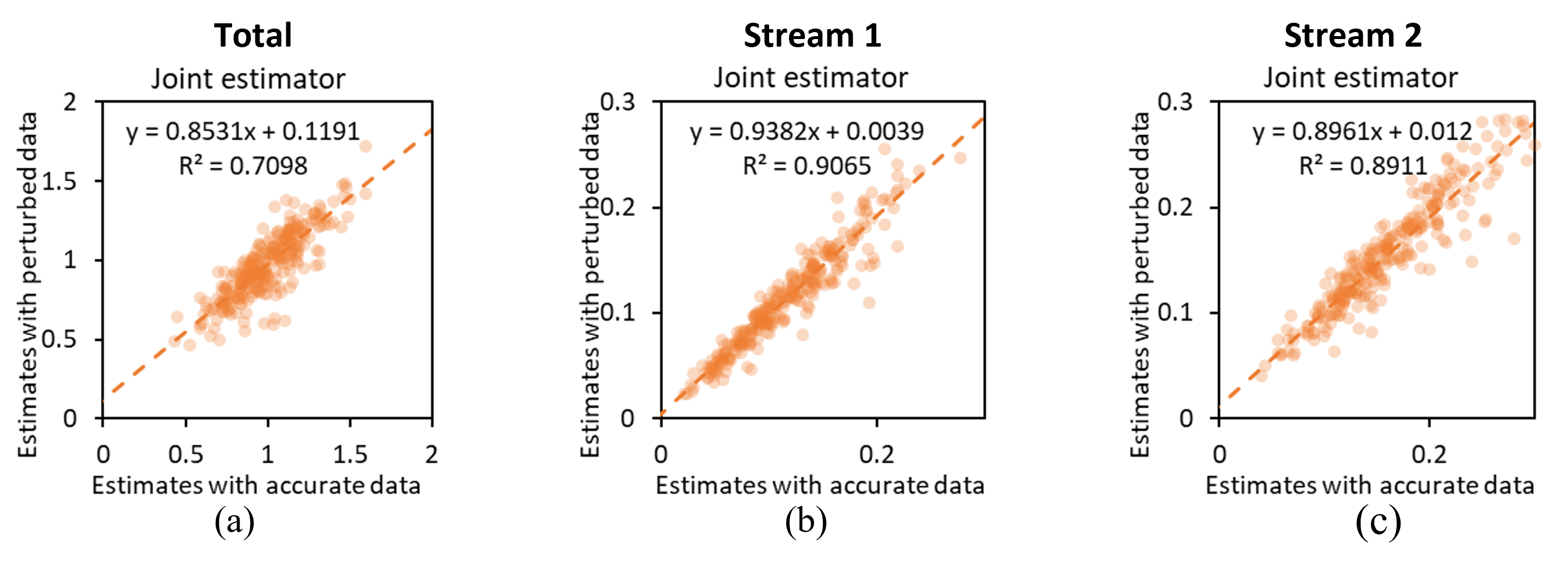}
	\caption{The performance of two arrival rate estimators: (a) total arrival rate of the intersection; (b) stream 1 arrival rate; (c) stream 2 arrival rate.}
	\label{FIG:arrival rate}
\end{figure}

The estimation results of arrival rates of all cycles are presented in Fig. \ref{FIG:arrival rate}, where the penetration rate is 0.5. Stream 1 is a left-turning stream, and stream 2 is a through-moving stream. Each data point is an estimate based on the accurate aggregated information (horizontal axis) versus the estimate based on the perturbed aggregated information (vertical axis). 

As is shown, most data points of the joint estimator are distributed around the line $y = x$, indicating that the estimates based on the perturbed aggregated information are close to the estimates based on the accurate aggregated information, which demonstrates that the joint estimator is robust against the perturbation on aggregated data. This explains why the performance of privacy-LP is almost identical to that of LP under different penetration rates, as indicated by Fig. \ref{FIG:7}.

\section{Conclusions and Future Studies}  \label{Conclusions}
In this study, we propose a paradigm for privacy-preserving CV-based adaptive traffic signal control that ensures the privacy of CV data with minimum impact on the control performance. This is one of the pioneering works that integrate privacy-preserving mechanisms into traffic signal control. The proposed method builds on a privacy-preserving data aggregation mechanism, which provides triple protections to CV data, including the use of key parameters, SMPC, and DP guarantees. Based on the outputs of the data aggregation mechanism, we formulate traffic state estimation and traffic signal control in a way that can effectively integrate the outcomes of the privacy-preserving data aggregation mechanism. Specifically, we formulate a multi-stream estimation model to estimate arrival flow to the intersection, which has been shown to successfully mitigate the impact of noises brought by the LAP mechanism. We formulate the traffic signal control problem into a simple but effective LP that takes the arrival flow and parameters calculated from the data aggregation mechanism as input, which is then extended to a stochastic programming problem to explicitly handle noises generated to ensure DP. 

We perform case studies of a typical isolated signalized intersection to evaluate the performance of the proposed models.  
Results show that both the deterministic and stochastic LP models outperform the actuated signal controller, demonstrating the effectiveness of CV-based signal control. It is further important to note that the resulting delay of the deterministic LP model is not sensitive to the noises generated to ensure DP, showing that adding DP noises can improve privacy with minimum impact on control performance. Finally, by comparing the stochastic and deterministic LP models, we can see that the stochastic LP model successfully handles noises in parameters, resulting in fewer residual vehicles and fewer stops, with barely any increase in vehicle delay. Overall, the proposed privacy-preserving traffic signal control algorithms are promising, as they can protect privacy with minimum impact on control performance.

This research opens several interesting directions for future work. First, in addition to the simulation-based validation in this paper, we would like to theoretically quantify the privacy-utility tradeoff in traffic signal control problems. Second, we will generalize the signal control algorithms for isolated intersections to coordinated intersections for urban arterial or urban networks, whereby sensitive traveler routing information is protected. Third, it would be interesting to extend the proposed methods to account for scenarios with more types of vehicles that request more intricate control actions, such as automated vehicles (AVs) that allow the coordination between signal control and vehicle control, as well as transit vehicles/emergency vehicles that request priority.

\bibliographystyle{unsrt}  
\bibliography{manuscript_0510}  

\appendix
\section{List of important variables}  \label{Appendix A}

\begin{longtable}[htbp]{l|p{14cm}}
\caption{List of important variables} \label{tab:nomenclature}\\
\hline\hline
\multicolumn{2}{c}{The following variables are for the privacy-preserving data aggregation mechanism} \\\hline
$\mathcal{K}$ & set of all controlled streams of the studied intersection, indexed by $k$. \\\hline
$\mathcal{I}$      & set of CVs in the zone of interest, indexed by $i$. \\\hline
$\mathcal{I}_k^{\rm{q}}$ & set of queued CVs in stream $k$ during red time. \\\hline 
$N$     &  number of CVs in $\mathcal{I}$, i.e., $N=|\mathcal{I}|$. \\\hline
$N_{k,r}^{\rm{q}}$ & number of queued CVs in stream $k$ at the $r$-th decision step in the past, where $r=1$ represents the current number of queued CVs in stream $k$.  \\\hline
$\mathcal{D}$ &  all possible datasets, indexed by $D,~D'$. \\\hline
$\phi: \mathcal{D} \rightarrow \mathcal{X}$ & query function, where $\mathcal{X}$ is the set of outputs.  \\\hline
$\mathcal{A}: \mathcal{D} \rightarrow \mathcal{X}$ & randomized mechanism in DP. \\\hline
$\Delta_{\phi}$ & global sensitivity of the function $\phi$. \\\hline
$\epsilon$ & privacy budget of differential privacy (DP). \\\hline
$\xi$ &  random variable generated by Laplace distribution $Lap(0,b)$ with the scaling parameter $b=\Delta_{\phi}/\epsilon$. $\xi_i$ represents the noise added by the Laplace mechanism (LAP) on $CV_i$.  \\\hline
$\beta$ & random variable generated by Beta distribution $Beta(1,N-1)$. \\\hline 
$\alpha_{k,i}$ & indicator of whether $CV_i$ is in stream $k$.  \\\hline
$\delta_i$ & indicator of whether $CV_i$ is in the queue before the stopline.  \\\hline
$p_i$ & current position of $CV_i$ measured from the stopline, calculated as $L_i/L_0$, where $L_i$ represents the current distance of the $CV_i$ measured from the stopline, and $L_0$ represents the jam spacing headway for queued vehicles.  \\\hline
$t_i$ & (expected) arrival time of $CV_i$ to the stopline (relative to the start time of the red signal).  \\\hline
$\eta_{k,i}$ & counting variable indicating whether $CV_i$ is a queued vehicle in stream $k$, i.e., $\eta_{k,i}=\alpha_{k,i}\delta_i$. The sum $\eta_k = \sum_{i\in\mathcal{I}}\eta_{k,i}$ represents the total number of queued CVs in stream $k$. \\\hline 
$P_{k,i}$ & position variable of $CV_i$ in stream $k$, calculated as $P_{k,i}=\alpha_{k,i} \delta_i p_i$. The sum $P_k = \sum_{i\in\mathcal{I}}P_{k,i}$ is used to calculate the arrival flow.  \\\hline 
$T_{k,i}$ & arrival-time variable of $CV_i$ in stream $k$, calculated as $T_{k,i}=\alpha_{k,i} \delta_i t_i$. The sum $T_k = \sum_{i\in\mathcal{I}}T_{k,i}$ is used to calculate the arrival flow.  \\\hline 
$x_i$ & general representation of private data $CV_i$ wants to share, where $x_i \in \{\eta_{k,i}, P_{k,i}, T_{k,i} \}_{k\in \mathcal{K}}$. The sum $x=\sum_{i\in\mathcal{I}}x_i$ represents the aggregated value that CVs in $\mathcal{I}$ want to calculate. \\\hline 
$s_{i,j}$ &  the $j$-th share that $CV_i$ sends to $CV_j$. \\\hline 
$S_i$ & general representation of the data $CV_i$ received from all CVs, i.e., $S_j=\sum_{i\in\mathcal{I}}s_{ij}$. Specifically, $S_{k,i}^\eta$, $S_{k,i}^P$, and $S_{k,i}^T$ represent the counting variable, position variable, and arrival-time variable $CV_i$ receives in relation to stream $k$.  \\\hline  
$\hat{S}_i$ & LAP perturbed data $\hat{S}_i = S_i+\xi_i$ that $CV_i$ submits to CV-DC. Specifically, $\hat{S}_{k,i}^\eta$, $\hat{S}_{k,i}^P$, and $\hat{S}_{k,i}^T$ represent the perturbed counting variable, position variable, and arrival-time variable $CV_i$ submits to CV-DC in relation to stream $k$. \\\hline
$\hat{\eta}_k$, $\hat{P}_k$, $\hat{T}_k$ & total vehicle count, total position variable, and total arrival-time variable in stream $k$. These data will be calculated by CV-DC and sent to TSCC for traffic signal control. \\\hline
\multicolumn{2}{c}{The following variables are for privacy-preserving traffic signal control} \\\hline
$\lambda_k$ &  arrival rate to stream $k$, which can be calculated using a function $F_{\lambda}(\cdot)$. \\\hline
$\gamma_k$ & proportion of CVs in stream $k$ over CVs in all streams. \\\hline
$g_k^s$ & green start time of stream $k$ (\textbf{decision variable}). \\\hline
$g_k^e$ & green end time of stream $k$ (\textbf{decision variable}). \\\hline
$g_k^{min}, g_k^{max}$ & minimum and maximum green time of stream $k$. \\\hline
$C$ & cycle length (\textbf{decision variable}), bounded by $C_{\min}\leq C \leq C_{\max}$. \\\hline
$\theta$ & set of decision variables, i.e., $\theta = \{C, \{g_k^s,g_k^e \}_{k \in \mathcal{K}} \}$. \\\hline
$\tau_i$ & expected stopline through time of $CV_i$. \\\hline 
$d_i$ & expected delay of a queued $CV_i$, i.e., $d_i=t_i-\tau_i$. \\\hline 
$l_k^s$ & start-up lost time of stream $k$. \\\hline 
$l_k^y$ & yellow lost time of stream $k$. \\\hline 
$h_k$ & average queue dissipation time headway of stream $k$ at the stopline. \\\hline 
$y_k$ & yellow time of stream $k$. \\\hline  
$a_k$ & red clearance time. \\\hline   
$r_k^{s,0}$ & red start time of stream $k$. \\\hline  
$Q_k$ & number of residual queued vehicles of stream $k$ (\textbf{decision variable} for the second stage problem of stochastic programming). \\\hline  
$\omega$ & random sample in the sample space $\Omega$. \\\hline  
$\mathcal{M}$ & set of scenarios in P2. \\\hline 
$Q_k^m$ &  number of residual queued vehicles of stream $k$ of scenario $m\in\mathcal{M}$ (\textbf{decision variable} for P2). \\\hline  
$P_k^m,T_k^m$ & parameters of scenario $m\in\mathcal{M}$ in P2. \\\hline 
\multicolumn{2}{c}{The following variables are for the determination of DP parameters} \\\hline
$P_{risk}$ & probability of a particular CV being identified in the database. \\\hline 
$P_{dire}$ & probability of a CV being identified in a specific direction. \\\hline 
$L_{link}$ & length of a link if the link spillback is not considered. \\\hline 
$\alpha^r$ & proportion parameter in DP for the vast majority. \\\hline 
$\Delta_{f,vast}$ & sensitivity for vast majority. \\\hline 
$Q_e$ & an empirically determined value to determine the sensitivity of $P_k$. \\\hline  
$\varphi$ & a coefficient greater than 0 to control the sensitivity of $T_k$. \\\hline 

\hline\hline
\end{longtable}

\section{Derivation process of arrival rate estimator}  \label{Appendix B}
\setcounter{equation}{0}
\renewcommand\theequation{B.\arabic{equation}}
Each stream $k \in \mathcal{K}$ has observed a set of queued CVs $\mathcal{I}_k^{\rm{q}}$ during the corresponding red time. Recall that $p_i$ and $t_i$ indicate the queuing position and expected arrival time of queued CV $i \in \mathcal{I}_k^{\rm{q}}$, respectively, and $\gamma_k$ denotes the proportion of CVs of stream $k$ (lane average) accounting for all streams over a historical period.

Assuming that vehicle arrival follows a Poisson process. Then, given the arrival rate $\lambda_k$, we can represent the distribution of the queued position $p_i$ and the expected arrival time $t_i$ for queued $CV_i$ arriving during the red time of stream $k$ as:
\begin{eqnarray}
    Pr(p_i,t_i | \lambda_k)=\frac{{(\lambda_k t_i)}^{p_i}}{p_i!} e^{-\lambda_k t_i}.
\end{eqnarray}     

Hence, for all queued CVs in stream $k$, we have the following likelihood function of the arrival rate of stream $k$ (assuming that the arrival of these CVs is independent): 
\begin{eqnarray}
    \mathcal{L}(\lambda_k |\boldsymbol{p_k},\boldsymbol{t_k})=Pr(\boldsymbol{p_k},\boldsymbol{t_k} | \lambda_k)=\prod_{i \in \mathcal{I}_{k}^{\rm{q}}} \frac{{(\lambda_k t_i)}^{p_i}}{p_i!} e^{-\lambda_k t_i},
\end{eqnarray}     
where $\boldsymbol{p_k}$ is the vector of queuing position of queued CVs in stream $k$; $\boldsymbol{t_k}$ is the corresponding vector of expected arrival time (relative to the red start moment).

To jointly model arrival rates of multiple streams, here we introduce the total arrival rate $\lambda_0$ for all streams. Then for each stream, we have $ \lambda_k=\lambda_0 \gamma_k,~k\in\mathcal{K} $, where the turning probabilities $\gamma_k$ is obtained \emph{a priori}.  
Here we treat each queued CV as an observation of $\lambda_0$. Then given all queued CVs in all streams, we have the following joint likelihood function of $\lambda_0$
\begin{eqnarray}
    \mathcal{L}(\lambda_0 |\boldsymbol{p_k},\boldsymbol{t_k},k \in \mathcal{K})=Pr(\boldsymbol{p_k},\boldsymbol{t_k},k \in \mathcal{K} | \lambda_0)=\prod_{k \in \mathcal{K}} \prod_{i \in \mathcal{I}_{k}^{\rm{q}}} \frac{{(\lambda_0 \gamma_k t_i)}^{p_i}}{p_i!} e^{-\lambda_0 \gamma_k t_i}.
\end{eqnarray}

Let $\partial \mathcal{L}(\lambda_0 |\boldsymbol{p_k},\boldsymbol{t_k},k \in \mathcal{K})/\partial \lambda_0 =0$, we can solve the joint estimator of $\lambda_0$ as
\begin{eqnarray}
    \lambda_0=\frac{\sum_{k \in \mathcal{K}} \sum_{i \in \mathcal{I}_{k}^{\rm{q}}} p_i}{\sum_{k \in \mathcal{K}} \gamma_k \sum_{i \in \mathcal{I}_{k}^{\rm{q}}} t_i}, \label{eq: b-total lamda}
\end{eqnarray}  
which, by $ \lambda_k=\lambda_0 \gamma_k$, gives 
\begin{eqnarray}
    \lambda_k=\gamma_k \frac{\sum_{k' \in \mathcal{K}} \sum_{i \in \mathcal{I}_{k'}^{\rm{q}}} p_i}{\sum_{k' \in \mathcal{K}} \gamma_{k'} \sum_{i \in \mathcal{I}_{k'}^{\rm{q}}} t_i}. \label{eq: lamda_k}
\end{eqnarray}

\end{document}